\documentclass[tightenlines,twoside,secnumarabic,preprintnumbers,notitlepage,onecolumn,floatfix,nofootinbib,11pt,prd]{revtex4-1}
\pdfoutput=1
\usepackage{ dsfont }
\usepackage[english]{babel}
\usepackage{amsmath,amssymb,bm,slashed}
\usepackage{graphicx,xcolor}
\usepackage[sort&compress]{natbib}
\usepackage{ctable}
\usepackage[normalem]{ulem}
\usepackage{hyperref}
\usepackage{cleveref}
\usepackage{subfigure} % for subfigures
\usepackage{lipsum}
\definecolor{red}{rgb}{1.0, 0, 0}
\definecolor{orange}{rgb}{1,0.,1}
 \usepackage{gensymb}
\allowdisplaybreaks

\bibpunct{[}{]}{,}{n}{}{,}

\newcommand{\ev}[1]{\ensuremath{\left\langle #1 %
                     \right\rangle}} % Expectation value
\newcommand{\tr}{\text{tr}}

\newcommand{\eV}{\,\mathrm{eV}}

\newcommand{\GeV}{\,\mathrm{GeV}}
\newcommand{\TeV}{\,\mathrm{TeV}}

\begin{document}

% =============================================================================
\title{Neutrino Anarchy and Renormalization Group Evolution}
\author{Vedran Brdar$^{1}$}            \email[Email: ]{vbrdar@uni-mainz.de}
\author{Matthias K\"onig$^{1}$}	       \email[Email: ]{m.koenig@uni-mainz.de}
\author{Joachim Kopp$^{1}$}            \email[Email: ]{jkopp@uni-mainz.de}
\affiliation{$^1$ PRISMA Cluster of Excellence and Mainz Institute for Theoretical Physics, \\
                  Johannes Gutenberg-Universit\"at Mainz, 55099 Mainz, Germany}
\date{\today} % FIXME
\preprint{MITP/15-104}
% =============================================================================

\begin{abstract}
  The observed pattern of neutrino mixing angles is in good agreement with the
  hypothesis of \emph{neutrino anarchy}, which posits that Nature has chosen
  the entries of the leptonic mixing matrix at random.  In this paper we
  investigate how stable this conclusion is under renormalization group
  effects.  Working in the simplest type-I seesaw model and two variants of the
  inverse seesaw model we study how the statistical distributions of the
  neutrino mixing parameters evolve between the Grand Unification scale and the
  electroweak scale.  Especially in the inverse seesaw case we find significant
  distortions: mixing angles tend to be smaller after RG running, and the Dirac
  CP phase tends to be closer to zero.  The $p$-value describing the
  compatibility between the observed mixing angles and the anarchy hypothesis
  increases by 10--20\%. This illustrates that RG effects are highly relevant
  for quantitative studies of the anarchy scenario.
\end{abstract}

%==============================================================================
\maketitle
%==============================================================================

%==============================================================================
\section{Introduction}
\label{sec:intro}
%==============================================================================

The flavor structure of leptons in the Standard Model (SM) is experimentally
well established, but our theoretical understanding of its origin is very poor.
The simplest postulate is that there is \emph{no} particular structure to the
neutrino mass matrix, but that its elements are randomly chosen
$\mathcal{O}(1)$ parameters. This scenario is called neutrino anarchy
\citep{Hall:1999sn, Haba:2000be}.  While one may view the idea of nature
choosing model parameters at random with some reservation, it does not seem so
unnatural given that nature also chooses the outcome of quantum mechanical
measurements at random.  Moreover, it can be considered a valid low-energy
description if the underlying UV-complete theory is sufficiently
complicated~\citep{Haba:2000be}.

Once we accept the idea of anarchy, it can be used to derive meaningful
predictions for the values of the neutrino mixing parameters.  In particular,
as shown in~\citep{Haba:2000be}, the requirement that different elements of
the mass matrices are statistically independent and that their distributions
are basis-independent leads uniquely to the conclusion that the lepton mixing
matrix $U_\text{PMNS}$ must have a flat distribution in the Haar measure of the
$U(3)$ group. This allows one to compute $p$-values for the experimentally
measured mixing parameters~\citep{Haba:2000be,deGouvea:2003xe,
Espinosa:2003qz,deGouvea:2012ac}. The original neutrino anarchy
scenario~\citep{Hall:1999sn} has been criticized in \citep{Espinosa:2003qz} on
the grounds that it cannot be formulated unambiguously if mass matrices that
only differ in their unphysical phases are considered identical.  This
viewpoint is valid if one assumes that nature does not ``know'' about the
unphysical phases at all, i.e.\ that they do not appear as hidden variables
in the theory. It becomes invalid if we accept the unphysical phases as
equipollent with the physical ones, for instance on the grounds that they might
have physical significance in an underlying UV-complete theory.

Of course the idea of anarchy is not restricted to the neutrino mass models in
which it was originally proposed, but has also been applied for instance to
scenarios with sterile neutrinos \cite{Heeck:2012zb,Gluza:2011nm},
models with extra dimensions \cite{Agashe:2008fe}, and in the context of
grand unified theories
(GUTs) \cite{Altarelli:2012ia,Altarelli:2002sg,Bennett:1988xi}. Moreover,
the idea that physical parameters are drawn from a statistical distribution
has received some attention in the context of the string landscape,
see for instance \cite{Donoghue:2005cf,Donoghue:2009me}.

In this paper we investigate how stable the predictions of neutrino anarchy are
under the effects of renormalization group (RG) running.  This is an important
question given that we do not know how and at what scale the random
selection of model parameters happens.  It could be a feature of a grand
unified theory, of quantum gravity, or it could be an outcome of the dynamics at
the end of inflation.  We will show that in the simplest type-I seesaw
scenario, which augments the SM by three heavy right-handed neutrinos, mixing
angles derived from anarchical mass matrices at a high energy scale can differ
from those at experimentally testable scales by several degrees.  The probability
of the neutrino oscillation parameters being at their experimentally observed
values under the anarchy hypothesis increases by $\sim 7\%$ because of
this. We show that this result does not depend on the statistical test
we choose to make the comparison.  We also investigate how our results change
in extended neutrino mass models. In particular, we show that renormalization
group effects become larger in inverse seesaw models.

The plan of the paper is as follows: In \cref{sec:anarchy}, we will review
neutrino anarchy and its predictions without including renormalization group
effects.  Comparing these predictions to data quantitatively, we will show that
results are robust with respect to the choice of statistical test.
\Cref{sec:rge} contains our main results: we
will investigate how the predictions of neutrino anarchy are modified when
renormalization group running from the GUT breaking scale
$M_\text{GUT} \sim 10^{16}$~GeV to the electroweak scale is
taken into account.  We will do this for the type-I seesaw scenario (\cref{sec:TypeI}),
for the phenomenological inverse seesaw model (\cref{sec:Inverse}), and for a
particular realization of the inverse seesaw model involving a gauged
$U(1)_{B-L}$ symmetry (\cref{sec:ISSBL}).  Besides
looking at the mixing angles and complex phases, we will also study
the predictions that neutrino anarchy makes for the
effective neutrino mass parameter probed in searches for neutrinoless double
beta ($0\nu2\beta$) decay. We will present our conclusions
in \cref{sec:conclusions}. In the appendices, we will offer details on the
renormalization group equations we employ. In particular for the
inverse seesaw model with $U(1)_{B-L}$ symmetry, these equations have,
to the best of our knowledge, not been presented before.

%==============================================================================
\section{Neutrino Anarchy}
\label{sec:anarchy}
%==============================================================================

%------------------------------------------------------------------------------
\subsection{Statistical tests}
\label{sec:statTests}
%------------------------------------------------------------------------------

We consider first the type-I seesaw scenario, where the SM is extended by
adding three right-handed singlet fermions $N_R^\alpha$. The relevant terms
in the Lagrangian are
\begin{align}
  \mathcal{L}_\text{type-I} \supset
    - Y_e^{\alpha\beta} \overline{e_R^\alpha} H^\dag L^\beta
    - Y_\nu^{\alpha\beta} \overline{N_R^\alpha} \tilde{H}^\dag L^\beta
    - \frac{1}{2} M^{\alpha\beta} \overline{N^\alpha} (N^\beta)^c
    + h.c. \,,
  \label{eq:L-mass-seesaw}
\end{align}
where $L^\alpha = (\nu_L^\alpha, e_L^\alpha)$
are the left-handed lepton doublets ($\alpha = e, \mu, \tau$),
$e_R^\alpha$ are the right-handed charged lepton fields, $H$ is the SM Higgs
doublet, and $\tilde{H} = i\sigma^2 H^*$ is the charge conjugate Higgs field.
The Yukawa matrices $Y_e$ and $Y_\nu$ are general complex $3 \times 3$ matrices,
while the Majorana mass matrix $M$ for the right-handed neutrinos is
a complex symmetric $3 \times 3$ matrix.
After electroweak symmetry breaking, the Higgs field acquires a vacuum expectation
value (vev) $\ev{H} = (0, v/\sqrt{2})$.
If the eigenvalues of $M$ are much larger than those of $m_D \equiv Y_\nu v/\sqrt{2}$,
the effective neutrino mass term at low energies becomes
\begin{align}
  \mathcal{L}_{m,\text{type-I}} \supset - \frac{1}{2} m_\nu^{\alpha\beta} \,
    \overline{(\nu_L^\alpha)^c} \nu_L^\beta + h.c.\,,
\end{align}
with
\begin{align}
  m_\nu = -v\left( \frac{v}{2} Y_\nu^T M^{-1} Y_\nu +\mathcal{O}\left( \frac{v^3}{M^3}\right)   \right)  \,.
\end{align}
The charged lepton mass matrix $m_e \equiv Y_e v/\sqrt{2}$
is diagonalized by a transformation
$e_L^\alpha \to V_e^{\alpha\beta} e_L^\beta$, $e_R^\alpha \to
V_e^{\prime\alpha\beta}
e_R^\beta$, and the neutrino mass matrix $m_\nu$ is
diagonalized by a transformation $\nu_L^\alpha \to V_\nu^{\alpha\beta}
\nu_L^\beta$.  The leptonic mixing matrix $U_\text{PMNS}$ is then given by $U_\text{PMNS}
= V_e^\dag V_\nu$.

The anarchy hypothesis states that the entries of $m_D$ and $M$ are randomly
and independently drawn from a statistical distribution.  One usually imposes
the additional requirement that this distribution is the same in any basis, i.e.\
it is invariant under transformations of the form $\nu^\alpha_L \to
U^{\alpha\beta} \nu^\beta$,
$N^\alpha \to U'^{\alpha\beta} N^\beta$, where $U$ and $U'$ are unitary matrices.  One
can show \citep{Haba:2000be} that basis independence implies that the
parameters of $U_\text{PMNS}$ have a flat distribution in the \emph{Haar
measure}~\citep{Haba:2000be}.  If the $U(3)$ matrix $U_\text{PMNS}$ is parameterized
in terms of the three mixing angles $\theta_{12}$, $\theta_{13}$, $\theta_{23}$,
the Dirac CP phase $\delta_\text{CP}$, the two Majorana CP phases $\phi_1$,
$\phi_2$, and the three unphysical phases $\delta_e$, $\delta_\mu$,
$\delta_\tau$ (see for instance \cite{Antusch:2003kp}),
the Haar measure reads
\begin{align}
  dU_\text{PMNS} = d(\sin^2 \theta_{12}) \, d(\sin^2 \theta_{23}) \, d(\cos^4 \theta_{13})
    \, d\delta_\text{CP} \, d\phi_1 \, d\phi_2 \, d\delta_e \, d\delta_\mu \, d\delta_\tau \,.
  \label{eq:Haar}
\end{align}
A flat distribution in the Haar measure thus means that the distributions of
the parameters $\sin^2 \theta_{12}$, $\sin^2 \theta_{23}$, $\cos^4
\theta_{13}$, $\delta_\text{CP}$, $\phi_1$, $\phi_2$, $\delta_e$, $\delta_\mu$,
and $\delta_\tau$ are
flat within their physically allowed ranges~\cite{deGouvea:2008nm}.
In order to obtain a Haar-flat leptonic mixing matrix, the elements
$Y_\nu^{\alpha\beta}$ of the neutrino Yukawa matrix must be drawn from
a Gaussian distribution $\propto \exp[-|Y_\nu^{\alpha\beta}|^2]$, and the
entries $M^{\alpha\beta}$ of the right-handed mass matrix must be drawn from a
Gaussian distribution $\propto
\exp[-|M^{\alpha\alpha}|^2/\mathcal{M}^2]$ for the diagonal elements and $\propto
\exp[-2 |M^{\alpha\beta}|^2/\mathcal{M}^2]$ for the off-diagonal
elements~\citep{Lu:2014cla}.  Here, $\mathcal{M}$ is a dimensionful
parameter that sets the overall mass scale of the right-handed neutrino.
This way of generating
random mass matrices will be used throughout this work.  The statistical
distribution of $\mathcal{M}$ does not follow from principles like basis independence and
statistical independence of parameters, unlike the distribution of
$|M^{\alpha\beta}|/\mathcal{M}$.
Therefore, $\mathcal{M}$ must be considered an ad hoc choice, and
consequently neutrino anarchy cannot make firm predictions for dimensionful
quantities like the neutrino mass squared differences.

A suitable statistical test for quantifying the compatibility of the data with
neutrino anarchy is the multidimensional Kolmogorov--Smirnov (KS)
test~\cite{deGouvea:2003xe, deGouvea:2012ac}.
For a given oscillation parameter $X$, the test statistic
\begin{align}
  D_\text{KS}(x_0) \equiv \sup_{x'} \big|\Theta(x' - x_0) - F(x') \big| \,,
  \label{eq:D-KS}
\end{align}
measures the maximum distance between the theoretically predicted cumulative
distribution function (CDF) $F(x')$ of $x$ and the experimentally determined
approximation $\Theta(x' - x_0)$ to the CDF. Since we have only one Universe to
measure the neutrino oscillation parameters in, we have to contend ourselves
with just one measurement $x_0$, therefore the experimental approximation to
the CDF is simply a Heaviside step function.  In our case, $X$ is one of the
parameters appearing in the Haar measure \eqref{eq:Haar}, namely $\sin^2
\theta_{12}$, $\sin^2 \theta_{23}$ or $\cos^4 \theta_{13}$. Since the predicted
distributions of these parameters are flat, we have $F(x') = x'$.  Thus, the test
statistic is simply given by $D_\text{KS}(x_0) = \frac{1}{2} + \big|\frac{1}{2} -
x_0\big|$.  The $p$-value
\begin{align}
  p_\text{KS}(x_0) \equiv 2 [1 - D_\text{KS}(x_0)] = 1 - |1 - 2 x_0|
  \label{eq:p-KS}
\end{align}
gives the probability that, for a randomly drawn value $x'$ of $X$, the
test statistic $D_\text{KS}(x')$ is larger than for the reference value $x_0$,
i.e.\ that $D_\text{KS}(x') \geq D_\text{KS}(x_0)$.  In other words, $p_\text{KS}(x_0)$
measures the probability that the parameter takes a value more ``extreme''
(closer to the minimum and maximum values of 0 and 1) than the reference value
$x_0$.  Small $p_\text{KS}$ means that $x_0$ is an unlikely outcome of the random
draw, i.e.\ that the data does not support the anarchy hypothesis.  For testing
the compatibility of the anarchy hypothesis with not only the measurement of
one parameter, but with measurements of all three mixing angles, we define the
probability of the data as
\begin{multline}
  P_\text{KS} \equiv
    \int_0^1 \! d\sin^2{\theta_{12}}\ d\sin^2{\theta_{23}}\ d\cos^4{\theta_{13}} \,
    \Theta\Big[
      p_\text{KS}(\sin^2\theta_{12}^\text{obs}) \,
      p_\text{KS}(\sin^2\theta_{23}^\text{obs}) \,
      p_\text{KS}(\cos^4\theta_{13}^\text{obs}) \\
    - p_\text{KS}(\sin^2\theta_{12}) \,
      p_\text{KS}(\sin^2\theta_{23}) \,
      p_\text{KS}(\cos^4\theta_{13})
    \Big] \,.
  \label{eq:P-KS}
\end{multline}
Here, $\theta_{ij}^\text{obs}$ denotes the measured values of the oscillation
parameters.  Note that the statistical independence of the parameters appearing
in the Haar measure implies that we can easily generalize this procedure to
include additional parameters, for instance the Dirac CP phase $\delta_\text{CP}$.
Since the experimentally measured parameter values~\citep{Gonzalez-Garcia:2014bfa}
\begin{align}
  \sin^2\theta_{12} = 0.304
  \qquad
    \sin^2\theta_{23} = 
    \begin{cases}  
      0.452 \,\,(\text{NH})\\
      0.579 \,\,(\text{IH})
    \end{cases}
  \qquad
  \sin^2\theta_{13} = 0.0218 \,,
\end{align}
depend on whether the mass ordering is assumed to be normal (NH) or inverted (IH),
we define the mass hierarchy-weighted probability of the data under the anarchy
hypothesis
\begin{align}
  \tilde{P}_\text{KS} = p_\text{NH} P_\text{KS,NH} + p_\text{IH} P_\text{KS,IH}  \,.
  \label{eq:Ptilde-KS}
\end{align}
Here $p_{NH}$ ($p_{IH}$) denotes the probability of obtaining a normal (inverted)
mass hierarchy in the anarchy scenario, and $P_\text{KS,NH}$
($P_\text{KS,IH}$) is
the Kolmogorov--Smirnov probability of the normal hierarchy (inverted hierarchy)
best fit point, computed according to \cref{eq:P-KS}.  Note that
$\tilde{P}_\text{KS}$ is dominated by the first term since $p_\text{NH} \sim 0.95$.
When comparing neutrino anarchy predictions to data in the remainder of this paper,
we will always use these weighted $\tilde{P}_\text{KS}$ values.

\begin{figure}
  \centering
  \begin{tabular}{cc}
    \includegraphics[width=0.48\textwidth]{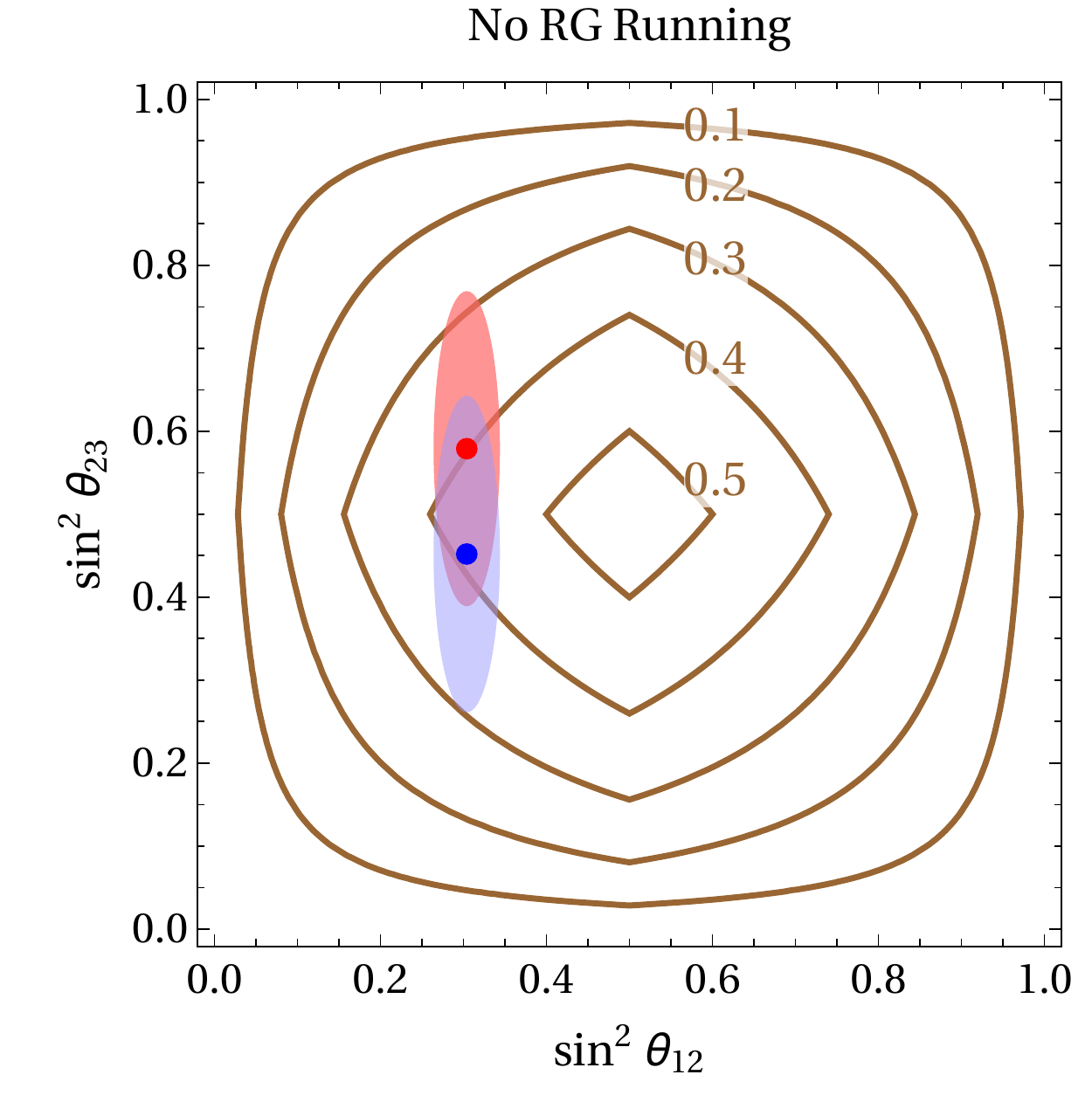} &
    \includegraphics[width=0.48\textwidth]{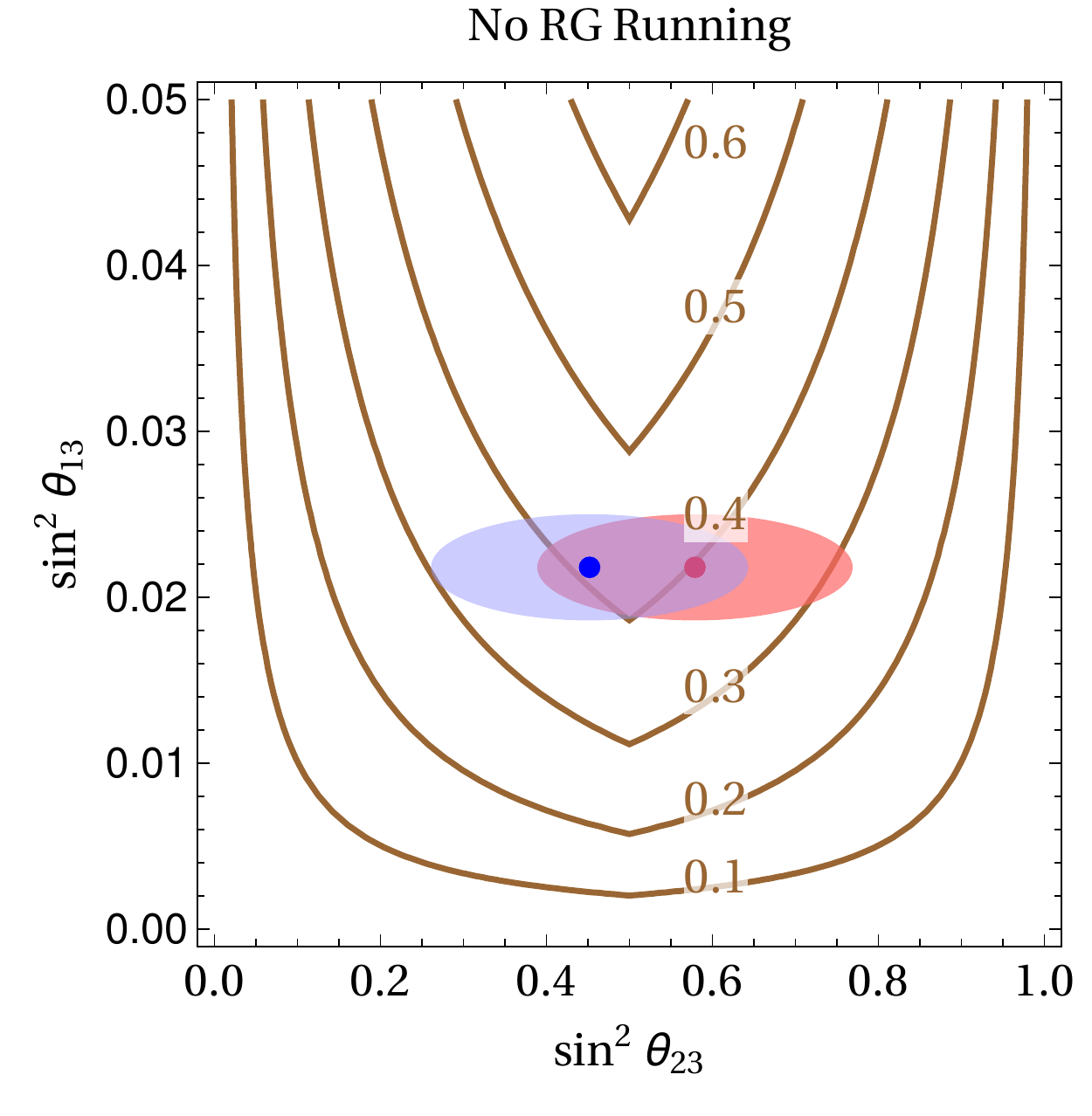} \\
    (a) & (b) 
  \end{tabular}
  \caption{(a) Contours of equal probability $P_\text{KS}$ (see
    eq.~\eqref{eq:P-KS}) in the $\sin^2 \theta_{12}$--$\sin^2 \theta_{23}$
    plane. The red (blue) ellipses show the favored region from the global
    fit in ref.~\citep{Gonzalez-Garcia:2014bfa} at the $3\sigma$ confidence
    level for inverted (normal) mass ordering. The colored dots indicate the
    corresponding best fit points. We have fixed the third
    mixing angle at its best fit value, $\sin^2 \theta_{13} = 0.0218$.
    (b) Contours of equal probability $P_\text{KS}$ in the
    $\sin^2 \theta_{23}$--$\sin^2 \theta_{13}$ plane, compared to the
    experimentally allowed regions for both mass orderings (same coloring as in (a)).
    We have fixed the solar mixing angle at its best fit value,
    $\sin^2 \theta_{12} = 0.304$.}
  \label{fig:KS-contours}
\end{figure}

Numerically, we find
\begin{align}
  \tilde{P}_\text{KS} = 0.411 \,.
  \label{eq:Ptilde-KS-number}
\end{align}
This implies that the probability that a set of mixing angles
randomly chosen according to the Haar measure gives worse $p$-values
than the experimentally observed ones is 41.1\%.
We can thus conclude that the data is well consistent with the anarchy
hypothesis, in accord with ref.~\citep{deGouvea:2012ac}.
\Cref{fig:KS-contours} shows the value of $P_\text{KS}$ as a function of
$\sin^2 \theta_{12}^\text{obs}$ and $\sin^2 \theta_{23}^\text{obs}$
(left panel) and of $\sin^2 \theta_{23}^\text{obs}$ and $\sin^2 \theta_{13}^\text{obs}$
(right panel), with the third mixing angle fixed in both cases
(see ref.~\citep{deGouvea:2012ac} for similar plots).

In addition to the KS test, we have considered two additional statistical tests,
namely the Anderson--Darling and the Cram\'er--von Mises tests.
For our simple case where only one measurement is available and parameters have
uniform distributions in the interval $[0,1]$,
the test statistic for the Anderson--Darling test is
\begin{align}
  D_\text{AD}(x_0) \equiv -1 - \log(1 - x_0) - \log(x_0) \,,
  \label{eq:D-AD}
\end{align}
and the one for the Cram\'er--von Mises test reads
\begin{align}
  D_\text{CvM}(x_0) \equiv \frac{1}{3} - x_0 + x_0^2 \,. 
  \label{eq:D-CvM}
\end{align}
It is, however, easy to see that the corresponding $p$-values
$p_\text{AD}(x_0)$ and $p_\text{CvM}(x_0)$, i.e.\ the probabilities
that a randomly drawn value $x'$ leads to a larger value for the test statistic
than the reference value $x_0$, are identical to the $p$-value for the
KS test, \cref{eq:p-KS}. Therefore, the Anderson--Darling and Cramer-von Mises
tests do not provide any additional insights. In this context, 
let us also mention ref.~\cite{Bergstrom:2014owa}, which
discusses a Bayesian comparison of various anarchical models to models
with flavor symmetries, finding consistence with both and a slight
preference for the latter. Similar conclusions have been reached also in
ref.~\cite{Altarelli:2012ia}.

Even though the predictions of the neutrino anarchy scenario depend on
the ad hoc choice of the right-handed mass scale $\mathcal{M}$, it is
possible to draw non-trivial conclusions on the relations between different
neutrino mass eigenvalues. A particularly interesting quantity to
consider in this context is the effective Majorana mass measured in
neutrinoless double beta decay,
\begin{align}
  \big| m_{ee} \bigr| \equiv \Big| \sum_{j} U_{ej}^2 m_j \Big| \,.
  \label{eq:m-eff}
\end{align}
where $U_{ej}$ are the elements in the first row of the leptonic mixing
matrix and $m_j$ are the light mass eigenvalues.
(In this paper, we always assume neutrinos to have Majorana masses,
so that neutrinoless double beta decay is in principle possible.)
We will therefore also discuss how neutrinoless double beta decay
experiments can probe the predictions of anarchy in the future, and how
these predictions are affected by renormalization group running.

%==============================================================================
\section{Renormalization Group Evolution and Anarchy}
\label{sec:rge}
%==============================================================================

Like all Lagrangian parameters in a quantum field theory, the neutrino masses and mixing
parameters are subject to renormalization group running.  It follows that, when
comparing the predicted probability distribution of a parameter to its
measured value, we have to be careful to evaluate both at the same
renormalization scale.  In particular, neutrino oscillation measurements are
typically carried out at energy scales of order MeV--GeV, while the flavor
structure of a typical seesaw model depends on scales $M_\text{GUT} \sim
10^{16}\GeV$. Therefore, we will in the following repeat the analysis of neutrino
anarchy, but now carefully taking into account renormalization group effects.

%------------------------------------------------------------------------------
\subsection{Type-I Seesaw Model}
\label{sec:TypeI}
%------------------------------------------------------------------------------

The first example we will study is the type-I seesaw scenario, which extends
the Standard Model by three heavy gauge singlet neutrinos with Majorana masses of 
$\mathcal{O}(10^{14}\GeV)$, which couple to the active SM neutrinos 
via a Yukawa coupling~\cite{Minkowski:1977sc, PhysRevLett.44.912}.
The Lagrangian is given by \cref{eq:L-mass-seesaw}.

The renormalization group equations for the type-I seesaw model have been
studied in great detail in
refs.~\cite{Antusch:2005gp,Mei:2005qp,Casas:1999tg,Chankowski:1993tx,Ohlsson:2013xva}. 
To investigate the effect of RG evolution on neutrino anarchy in this model, we
have randomly generated $10^{5}$ sets of parameters, defined by the Yukawa matrices $Y_\nu$
and the right-handed neutrino mass matrices $M$.
We assume these randomly generated matrices to enter the Lagrangian
at the high scale $M_\text{GUT}=10^{16}$~GeV.  As explained in ref.~\cite{Lu:2014cla},
to ensure basis invariance of the probability distributions and statistical
independence of different mass matrix elements, the elements of $Y_\nu^{\alpha\beta}$ must
follow a Gaussian distribution $\propto \exp[-|Y_\nu^{\alpha\beta}|^2/\mathcal{Y}]$.
The diagonal elements $M^{\alpha\alpha}$ of the right-handed Majorana mass
matrix follow a similar distribution $\propto \exp[-|M^{\alpha\alpha}|^2/\mathcal{M}]$,
while the distribution of the off-diagonal elements $M^{\alpha\beta}$ of $M$ is
$\propto \exp[-2 |M^{\alpha\beta}|^2/\mathcal{M}]$. Here $\mathcal{Y}$ is a dimensionless
parameter which we choose to be one, while $\mathcal{M}$ is dimensionful, and
we choose $\mathcal{M} = 10^{14}$~GeV to reproduce the correct order of magnitude
for the light neutrino masses. Below we will discuss the impact of different choices
for $\mathcal{M}$.

For the parameters of the standard model at the high scale we have chosen a set
of values offered by the \texttt{REAP} package~\cite{Antusch:2005gp}:
\begin{equation}
 \begin{aligned}
  Y_u(M_\text{GUT}) &=
    \begin{pmatrix}
      5.4039\cdot 10^{-6} &                     & \\
                          & 1.5637\cdot 10^{-3} & \\
                          &                     & 0.4829
    \end{pmatrix} \,, \\
  Y_d(M_\text{GUT}) &=
    10^{-5} \times
    \begin{pmatrix}
      2.1176            &  4.5697 + 0.0112 i &   2.1367 + 0.2581 i \\
      4.5697 - 0.0112 i & 22.3698            &  25.7097 - 0.0019 i \\
      2.1367 - 0.2571 i & 25.7097 + 0.0019 i & 614.0010
    \end{pmatrix} \,, \\
  Y_e(M_\text{GUT}) &=
    \begin{pmatrix}
      2.8370 \cdot 10^{-6} &                      & \\
                           & 0.5988 \cdot 10^{-3} & \\
                           &                      & 10.1789 \cdot 10^{-3}
    \end{pmatrix} \,.
  \end{aligned}
  \label{eq:ic}
\end{equation}

Of course the low-scale values of the quark and charged lepton Yukawa matrices
depend on the additional field content beyond the SM and on the values of the
new parameters introduced.  Therefore, we typically cannot reproduce the SM
Yukawa couplings exactly, with deviations being of order 10\%.
Nevertheless, since our interest here is in the running of the neutrino masses
and mixing parameters, these small deviations are negligible for our purposes.

We evolve each parameter point down to a lower scale, which we take to be
$M_Z=91.19\GeV$.  In principle, we should
evolve the mixing parameters down to an even lower scale, integrating out SM
particles along the way.  However, as we will see, the dominant contribution to
the running of the mixing parameters arises at scales $\gg M_Z$ and thus it
is justified to stop the RG evolution at $M_Z$. The renormalization group
equations for the type-I seesaw scenario have been derived in \cite{Antusch:2005gp},
and we follow the exact same procedure. Both at $M_\text{GUT}$ and at $M_Z$, we
extract the values of the physical mixing parameters by diagonalizing the
mass matrices and then following appendix~A of ref.~\cite{Antusch:2003kp}.
In diagonalizing $m_\nu$, we have to specify a convention for the ordering of
the mass eigenvalues. We always choose them to be in ascending order
(case~A3 in ref.~\cite{deGouvea:2008nm}).  Note that for the
running of the neutrino \emph{masses}, decoupling effects cannot be neglected
and even the top-quark decoupling may become important~\cite{Haba:2014uza}.
However, since we are mostly interested in the running of the
mixing angles and CP phases here, it is justified to
neglect decoupling effects. This is especially true for the top and the Higgs
because the running of the mixing parameters is negligible at their mass scale
(see e.g.\ \cref{fig:largeRunning} below).

\begin{figure}
  \centering
  \includegraphics{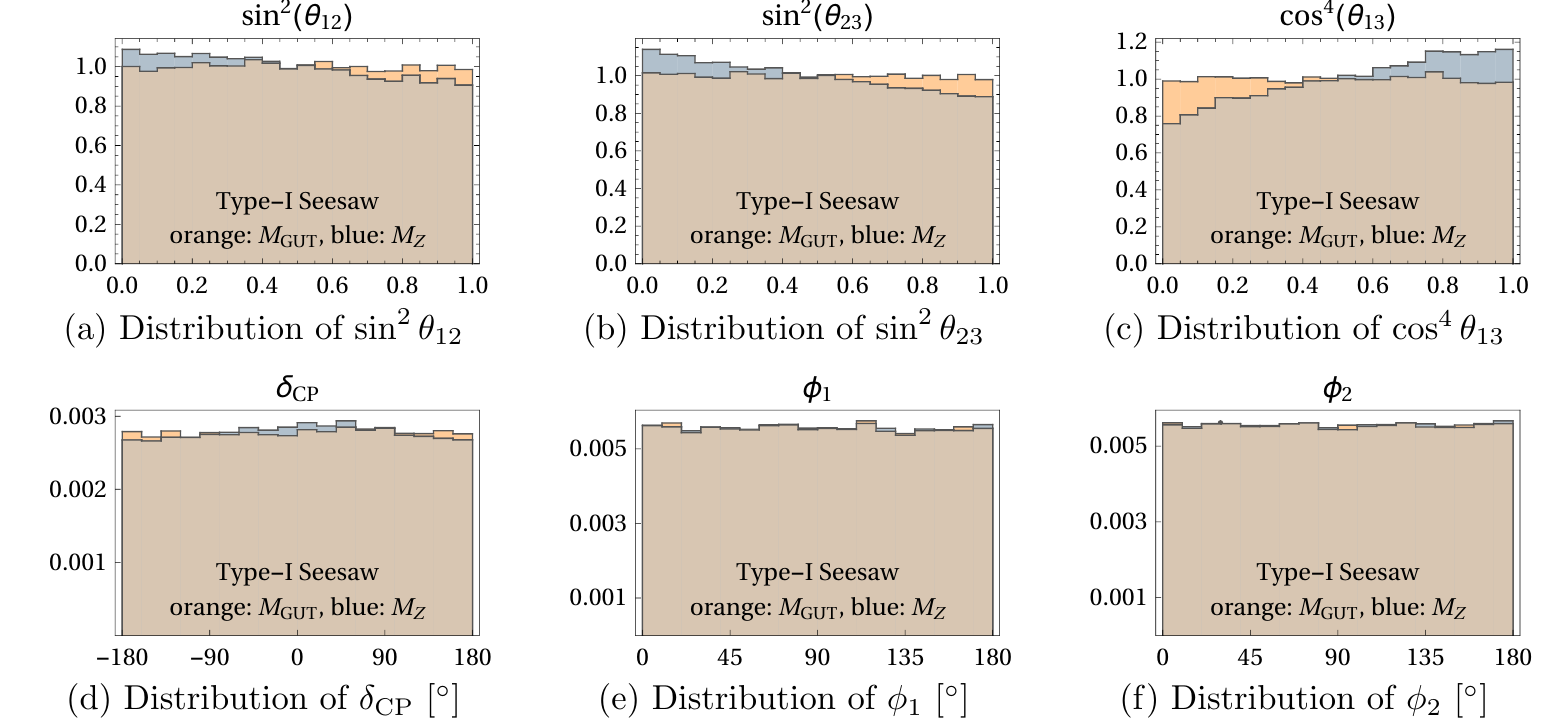}
  \centering
  \caption{Distributions of the three mixing angles, the Dirac CP phase
    $\delta_\text{CP}$ and the two Majorana CP phases $\phi_1$ and $\phi_2$
    before and after renormalization group running in the type-I seesaw model.
    Orange regions correspond to parameters at $M_\text{GUT}$, blue regions to
    parameters at $M_Z$.  Note that the physical range for $\delta_\text{CP}$
    is $[-180\degree, 180\degree)$, while the physical ranges for $\phi_1$ and
      $\phi_2$ are only $[0, 180\degree)$~\cite{deGouvea:2008nm}.}
    \label{fig:RGEDistSM}
\end{figure}

\begin{figure}
  \centering
\includegraphics{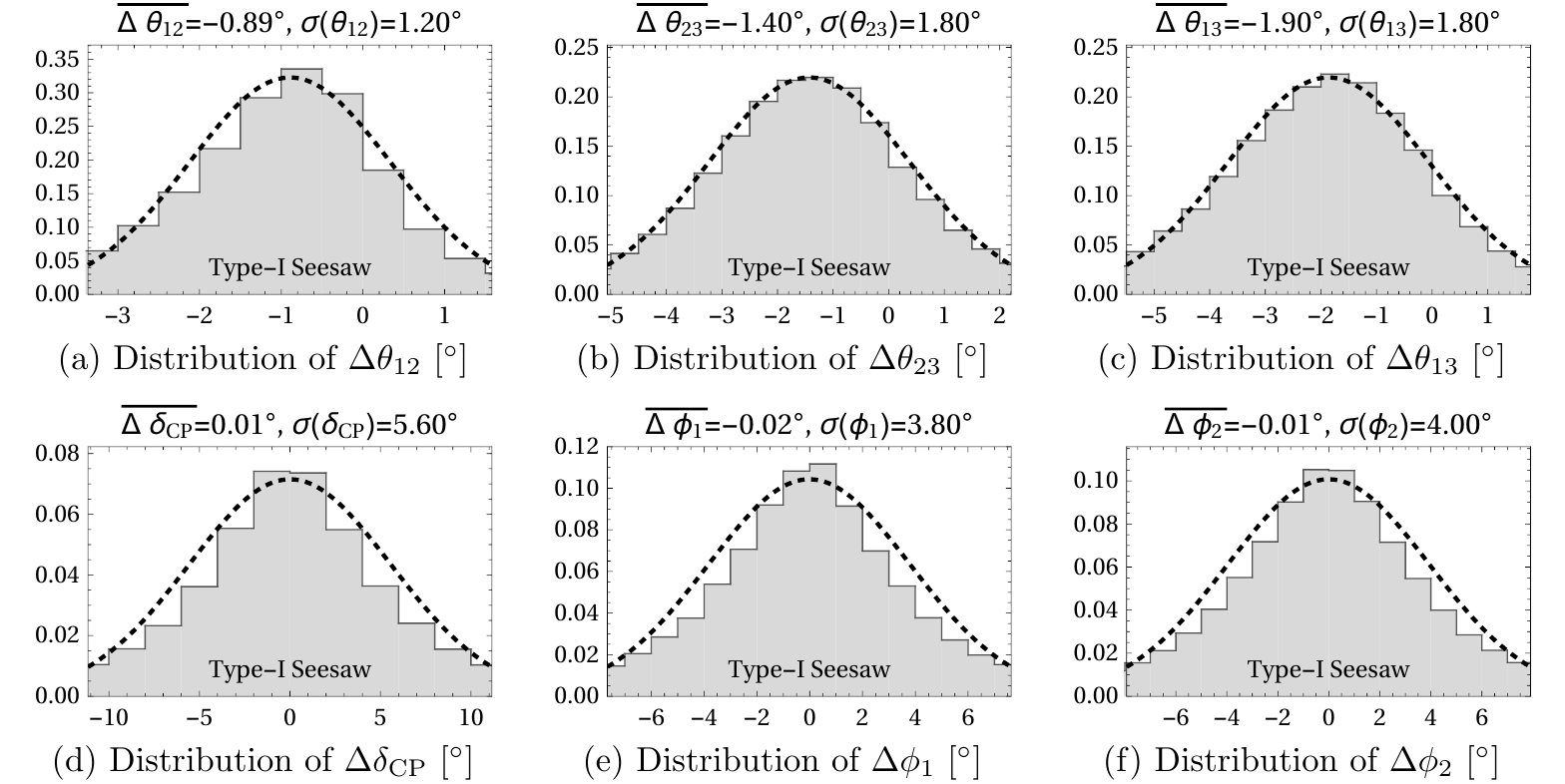}
  \caption{Distributions of the RG-induced shifts in the mixing parameters,
    $\Delta x \equiv x(M_Z) - x(M_\text{GUT})$ in the type-I seesaw model. Also
    shown are the central values $\overline{\Delta x}$
    and the widths $\sigma(x)$ of Gaussian fits
    to the distributions.}
  \label{fig:RGEDeltaSM}
\end{figure}

We show our results in \cref{fig:RGEDistSM}.  In each panel, we compare the
distribution of one of the mixing parameters at $M_\text{GUT}$ (orange
histograms) to the distribution at $M_Z$ after RG running (blue histograms).
By construction, the distributions at $M_\text{GUT}$ are flat, while at $M_Z$,
we observe a small bias towards smaller mixing angles.  We show in
\cref{fig:RGEDeltaSM} the distributions of the RG-induced shifts,
$\Delta x = x(M_Z) - x(M_\text{GUT})$, where $x \in \{ \theta_{12}, \theta_{13},
\theta_{23}, \delta_\text{CP}, \phi_1, \phi_2 \}$.  We have
also fitted each histogram in \cref{fig:RGEDeltaSM} to a Gaussian distribution 
\begin{align}
  f(x) = \frac{1}{\sqrt{2\pi} \sigma(x)}
         \exp\bigg[-\frac{(x-\overline{x})^2}{2[\sigma(x)]^2} \bigg] \,,
\end{align}
and we give the fitted values of $\overline{x}$ and $\sigma(x)$ in the plots.

\begin{figure}
 \begin{center}
   \includegraphics[width=0.5\textwidth]{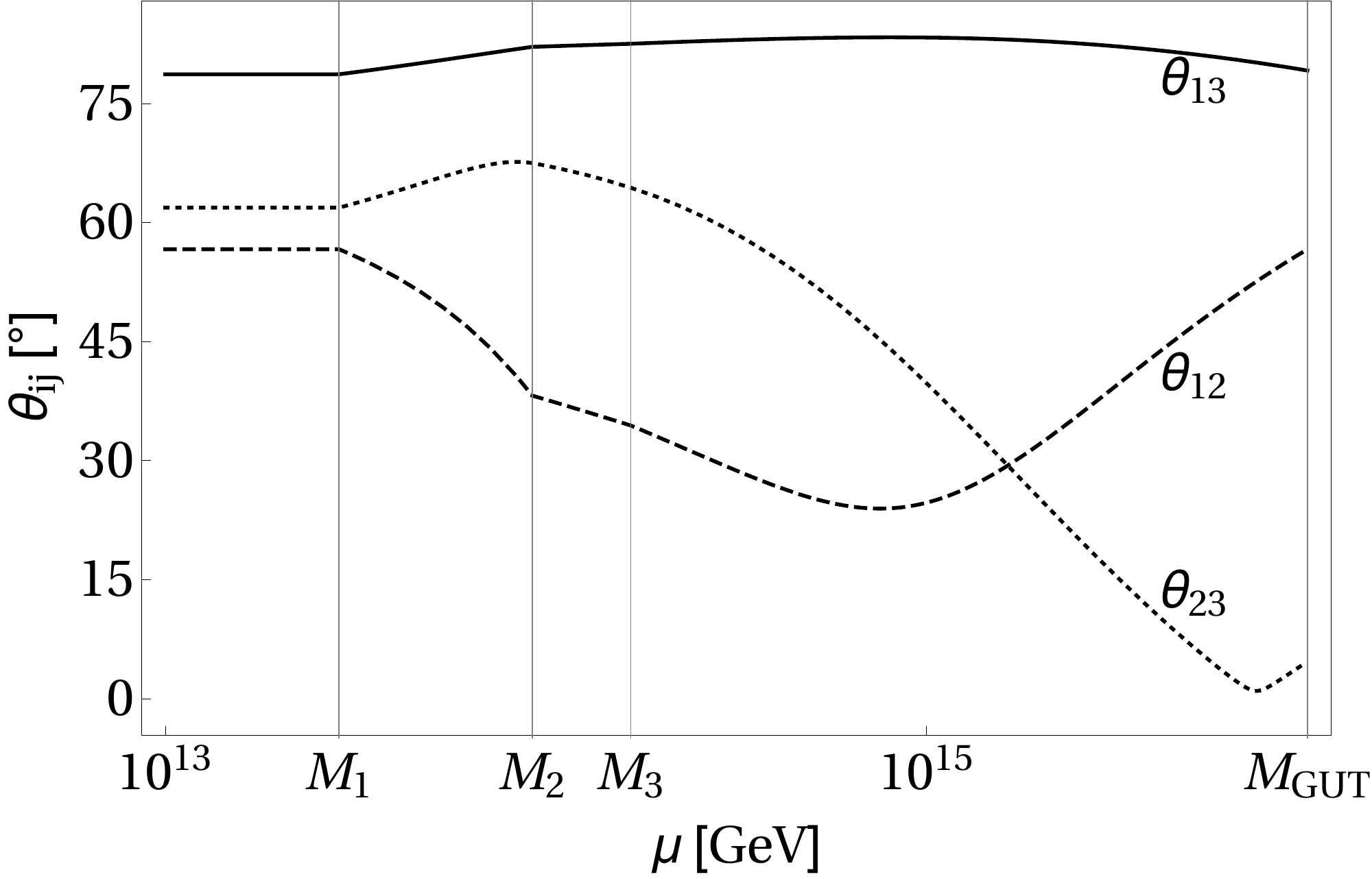}
 \end{center}
 \vspace{-0.5cm}
 \caption{A parameter point of the type-I seesaw model with large running effects
   in the neutrino mixing angles. The vertical dotted lines show the decoupling
   thresholds of the heavy singlet neutrinos, $M_1=2.9\cdot 10^{13}\GeV$,
   $M_2=9.2\cdot 10^{13}\GeV$ and $M_3=1.67\cdot 10^{14}\GeV$ . Note how RG
   running ceases after the last heavy neutrino is integrated out. The masses
   of the active neutrinos at the low scale are $m_1 = 0.02\eV$, $m_2=0.32\eV$
   and $m_3=0.36\eV$.}
 \label{fig:largeRunning}
\end{figure}

We find a slight preference for the mixing angles to decrease by 1--2$^\circ$
after evolving them from $M_\text{GUT}$ to $M_Z$: the distributions of
$\Delta\theta_{ij}$ all peak at small negative values.
The distribution of the
Dirac CP phase $\delta_\text{CP}$ in \cref{fig:RGEDistSM} (d) shows a
slight peak at zero after running,
while the distributions of the Majorana phases $\phi_1$ and $\phi_2$
in \cref{fig:RGEDistSM} (e), (f) are flat
both at $M_\text{GUT}$ and at $M_Z$. These are of course only
statements about the average behavior of a large set of random parameter
points---individual points can lead to much larger running effects. In
particular, for scenarios where at least two neutrino mass eigenvalues are very
close to each other, the running of the mixing angles is known to be
potentially large~\cite{Antusch:2005gp}.

Even for parameter points that exhibit significant RG running,
we observe that most of the running happens at
scales above the masses of the heavy singlet neutrinos. \Cref{fig:largeRunning}
shows an example where the mixing angles exhibit large RG running.  We observe
that the running gradually diminishes as the heavy singlet neutrinos are
integrated out and completely ceases below the mass of the lightest of them.
This observation allows us to draw conclusions on the impact of the right-handed
neutrino mass scale $\mathcal{M}$ on the RG evolution: For $\mathcal{M}$
larger (smaller) than our choice $\mathcal{M} = 10^{14}$~GeV,
the singlet neutrinos are typically heavier (lighter), so the running ceases
sooner (later), and the overall shift in the parameters will be smaller (larger).
We explicitly checked this behavior numerically.

The fact that the mixing angles run preferentially \emph{downwards} can
be understood in the following way. First, recall that the leptonic
mixing matrix depends on the matrix $V_\nu$ that diagonalizes $m_\nu$
\emph{and} on the matrix $V_e$ that diagonalizes $Y_e^\dag Y_e$
according to $V_e Y_e^\dag Y_e V_e^\dag = D$, where $D$ is diagonal.
The flavor non-trivial terms in the $\beta$ functions for the mass and
Yukawa matrices (see \cref{sec:iss-appendix}) are dominated by the
$\mathcal{O}(1)$ terms containing $Y_\nu$.
The subdominant terms involving $Y_e$ are negligible. Since $Y_\nu$
is randomly chosen, it cannot lead to a preference for one sign over the
other in the $\beta$ function. Thus, the only possible source of such
a preference can come from the initial conditions for $Y_e$.
The $\beta$ function for $Y_e$ reads
\begin{align}
  16 \pi^2 \beta_{Y_e} = -\frac{3}{2} Y_{e} \big( Y_\nu^\dagger Y_\nu \big) + \dots \,.
  \label{eq:Ye-rge}
\end{align}
where the dots indicate the subdominant and flavor-diagonal terms.
The initial condition $Y_e(M_\text{GUT})$ for $Y_e$ is a
diagonal matrix (see \cref{eq:ic}),
with its strongly hierarchical eigenvalues sorted in ascending
order. The structure of \cref{eq:Ye-rge} shows that after RG running, the
elements in the $k$-th row of $Y_e$ will change by an amount proportional to
the $(kk)$ elements of $Y_e(M_\text{GUT})$. To proceed, it is easiest to
consider a two flavor system for illustrative purposes.  In this case, using
the hierarchy of the elements of $Y_e$ after running, one can show that the
mixing angle parameterizing $V_e$ is positive.  Since $U_\text{PMNS} = V_e^\dag
V_\nu$, and since the mixing angle describing $V_\nu$ is random in $[-\pi,
\pi]$, the mixing angle describing $U_\text{PMNS}$ is then preferentially
negative. In models where the dominant term in \eqref{eq:Ye-rge} comes 
with a different sign, the preferred direction of the running would be
reversed. One example of such model is a type-II seesaw model \cite{Schmidt:2007nq}.

To study how RG running affects the compatibility of neutrino anarchy with
the observed mixing parameters, we use again the Kolmogorov--Smirnov test.
Note that the multidimensional KS test described in
\cref{sec:anarchy} is only meaningful if the parameters in the integral
\cref{eq:P-KS} are statistically independent and follow a flat distribution.
We have checked that, even after RG running, the parameters $\sin^2\theta_{12}$,
$\sin^2\theta_{23}$ and $\cos^4\theta_{13}$ are still nearly uncorrelated.
For instance, the modulus of Pearson's correlation coefficient, $\big| \ev{x \cdot y}
/ (\ev{x} \ev{y}) \big|$, is $\lesssim 0.02$ for all
combinations of $x,\, y = \sin^2\theta_{12},\,
\sin^2\theta_{23},\, \cos^4\theta_{13}$.
To obtain flat distributions, we define new parameters
$\sin^2\theta_{12}'$,
$\sin^2\theta_{23}'$ and $\cos^4\theta_{13}'$ in the following way:
we fit each of the distributions in \cref{fig:RGEDistSM} (a), (b), (c) by
a cubic polynomial $f(x)$, where $x = \sin^2\theta_{12}$,
$\sin^2\theta_{23}$, $\cos^4\theta_{13}$. We normalize $f(x)$ such
that its integral over the physical range of $x$ is 1. We then define
$x'(x) \equiv \int_0^{x} \! d\tilde{x} \, f(\tilde{x})$.
Replacing the unprimed variables $x$ in
\cref{eq:P-KS} by the primed ones $x'(x)$, the KS test can now be applied in
a meaningful way.

\begin{figure}
  \centering
  \begin{tabular}{cc}
    \includegraphics[width=0.48\textwidth]{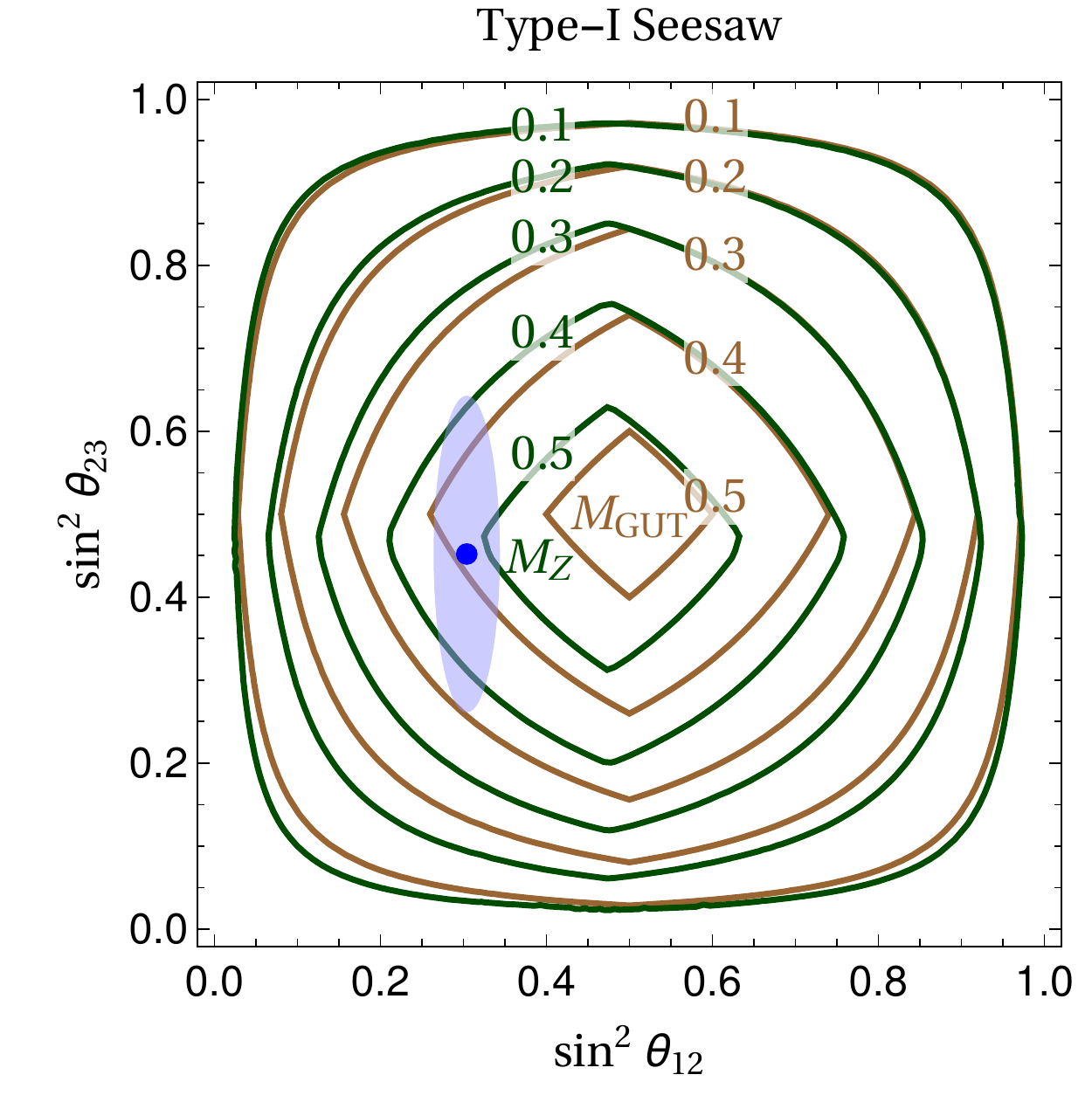} &
    \includegraphics[width=0.48\textwidth]{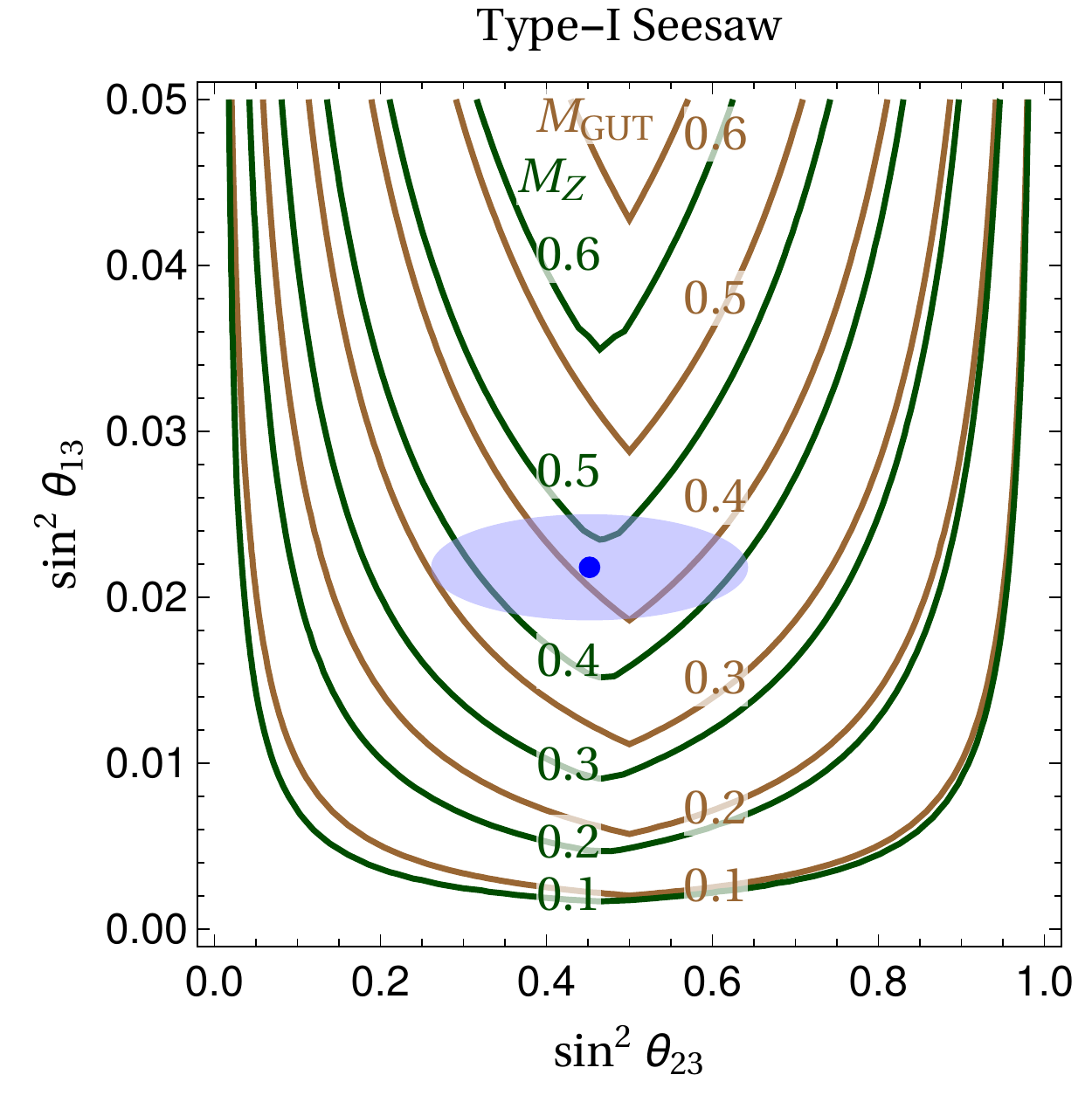} \\
    (a) & (b)
  \end{tabular}
  \caption{(a) Contours of equal probability $P_\text{KS,NH}$ (see
    \cref{eq:P-KS,eq:Ptilde-KS}) in the $\sin^2 \theta_{12}$--$\sin^2 \theta_{23}$
    plane with RG effects included (dark green) and without RG effects (brown, see
    \cref{fig:KS-contours} (a)) for the type-I seesaw model.
    We consider only parameter points with a normal mass ordering (see text
    for details).
    The blue ellipse indicates the normal ordering favored region from
    the global fit in ref.~\citep{Gonzalez-Garcia:2014bfa} at the $3\sigma$
    confidence level, and the blue dot is the corresponding best fit point.
    We have fixed the third mixing angle at its best fit
    value, $\sin^2 \theta_{13} = 0.0218$.  With RG effects included, the
    likelihood of the data with respect to anarchy increases from 41.1\%
    to 47.7\%. (b) Contours of equal probability $P_\text{KS,NH}$
    in the $\sin^2 \theta_{23}$--$\sin^2 \theta_{13}$
    plane with and without RG effects included (same coloring as in (a)).
    We have fixed the solar mixing angle at its best fit
    value, $\sin^2 \theta_{12} = 0.304$.}
  \label{fig:rgeVersusNoRGESM}
\end{figure}

A further subtlety arises because the RG evolution is different for parameter
points with a normal mass ordering and parameter points with an inverted
mass ordering.  We therefore separate our randomly generated points
according to the predicted mass ordering and compute $P_\text{KS,NH}$
($P_\text{KS,IH}$) in \cref{eq:Ptilde-KS} using only points with
normal (inverted) ordering. We find that the probability $\tilde{P}_\text{KS}$
of the observed parameter values, given the anarchy hypothesis,
increases by $\sim 7\%$ from 41.1\% to 47.7\% due to RG effects. Note that,
as before, $\tilde{P}_\text{KS}$ is entirely dominated by parameter points with a
normal mass ordering because the probability of obtaining an inverted ordering
from anarchy is only at the 5\% level.
We show in \cref{fig:rgeVersusNoRGESM}
the contours of equal Kolmogorov--Smirnov probability $P_\text{KS,NH}$
for parameter points with normal mass ordering before and after running.

Finally, we also comment on the running of the neutrino mass parameters, which
we find to be substantial. Parameterizing their approximate evolution  as
$dm/d\log\mu = -\gamma m$, we find for the anomalous dimension of the three
light mass eigenvalues an average value of $\gamma \simeq -0.027$.
This implies that the light neutrino masses run down by about 60\% between
$M_\text{GUT}$ and $M_Z$.

\begin{figure}
  \centering
  \begin{tabular}{cc}
    \includegraphics[width=.48\textwidth]{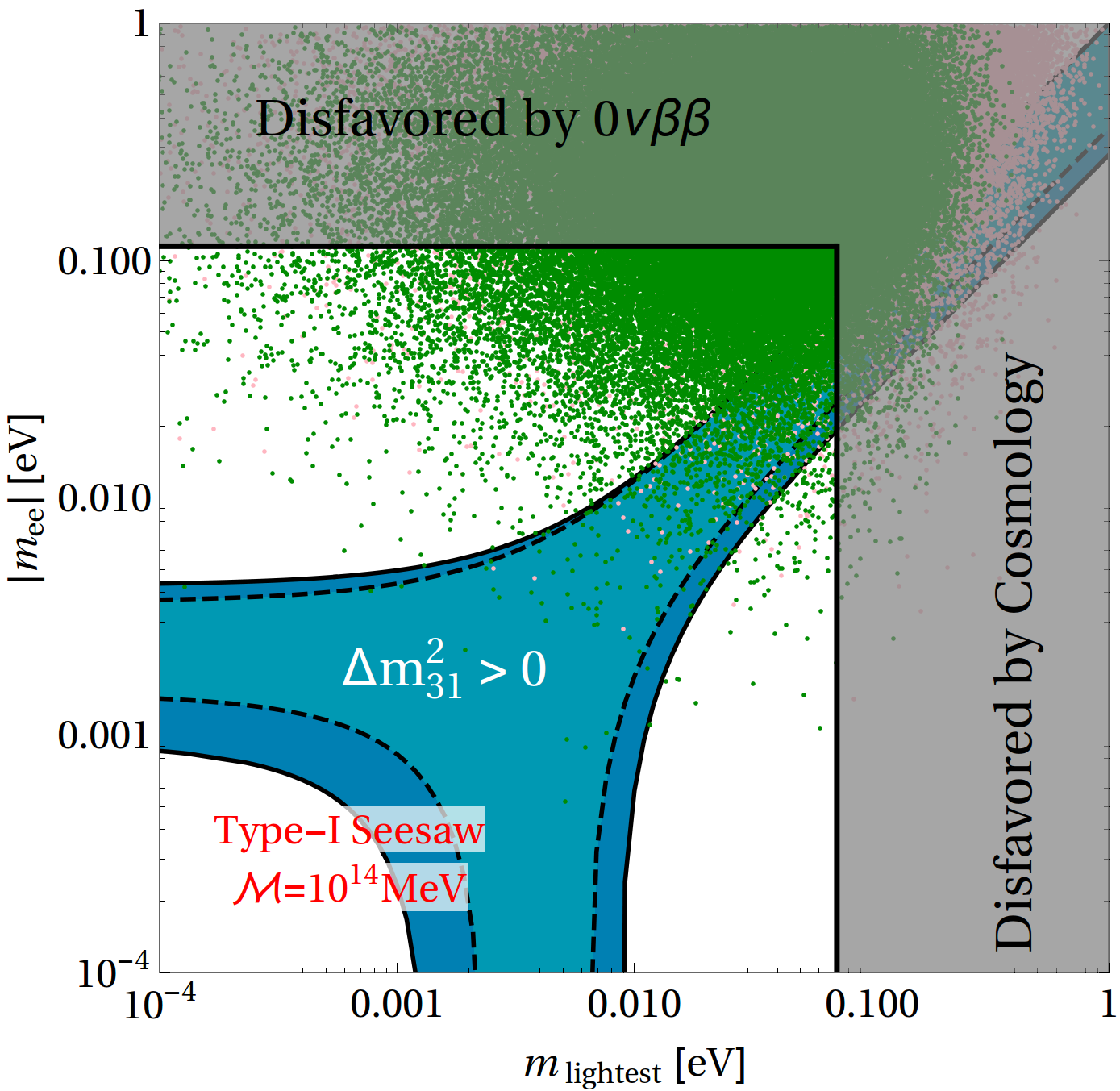} &
    \includegraphics[width=.48\textwidth]{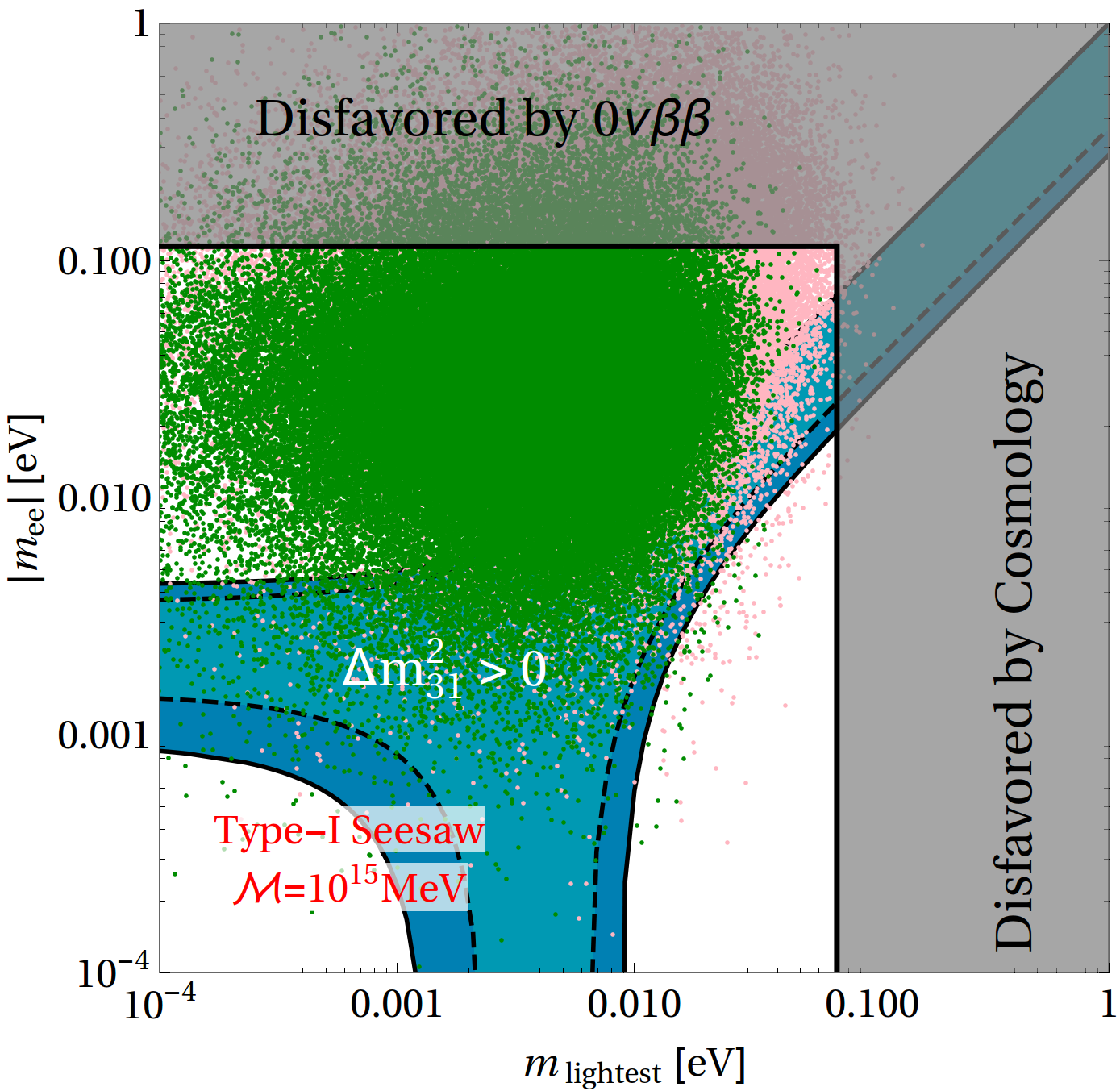} \\
    (a) & (b) \\[0.2cm]
    \includegraphics[width=.48\textwidth]{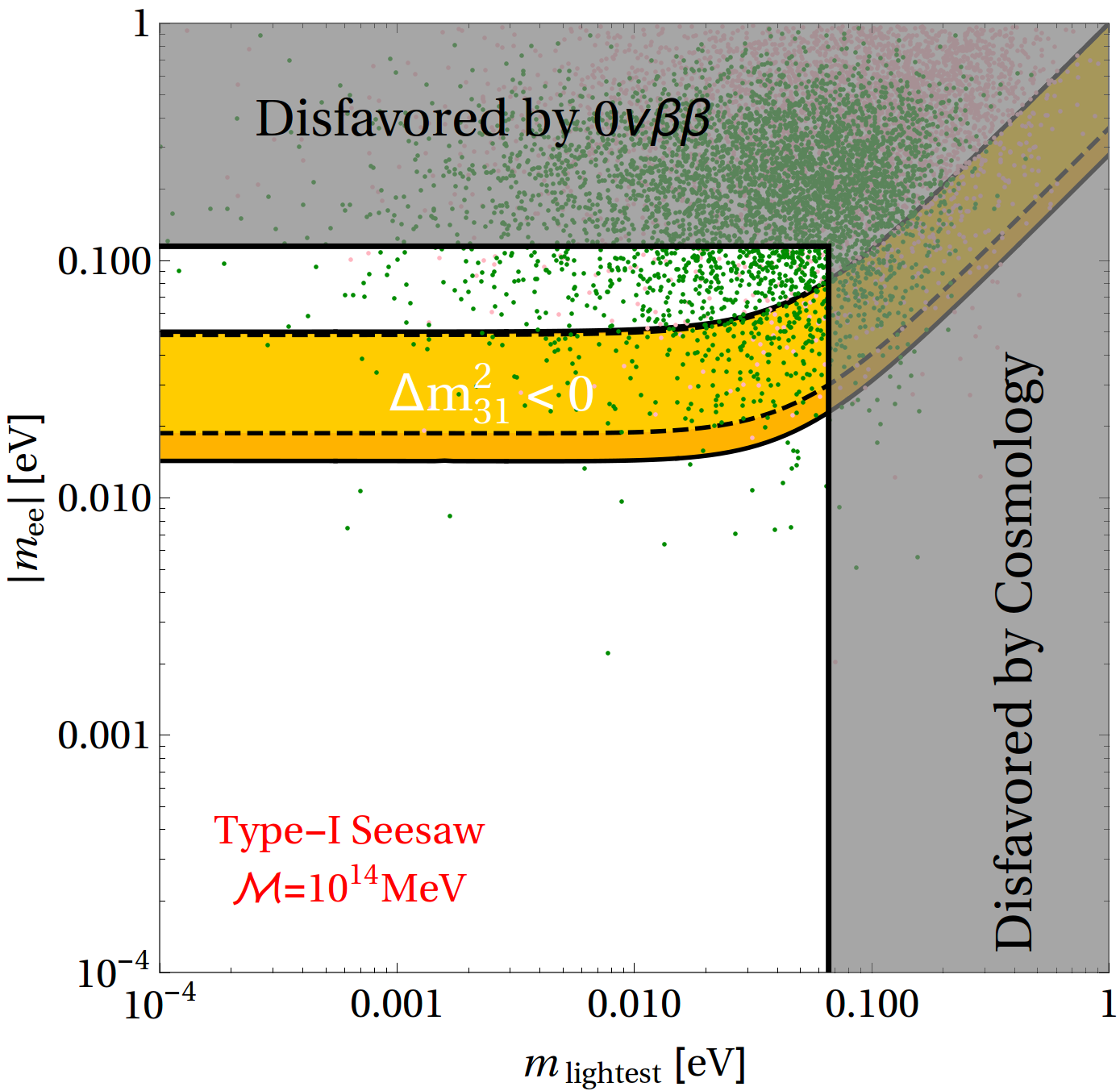} &
    \includegraphics[width=.48\textwidth]{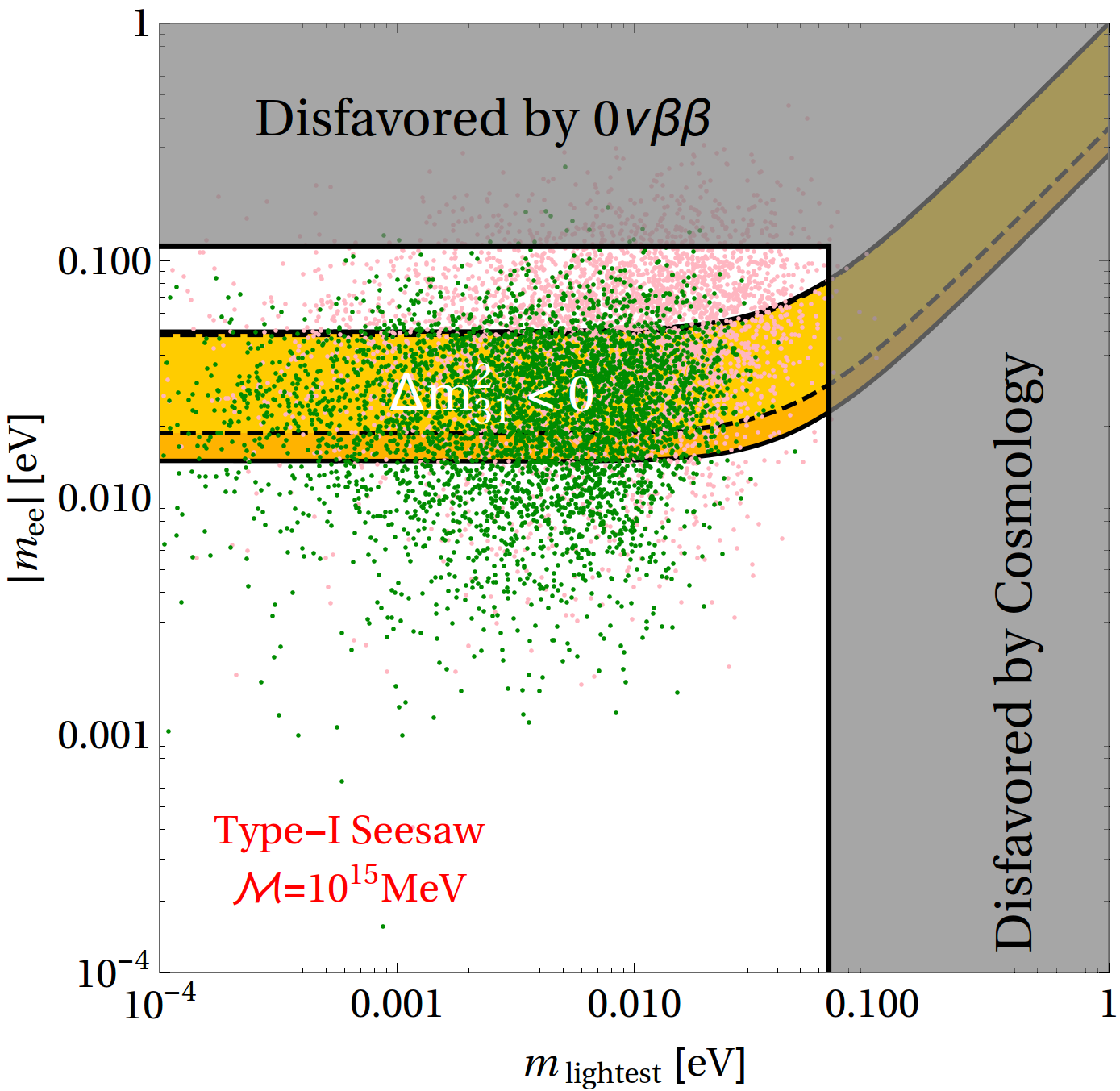} \\
    (c) & (d)
  \end{tabular}
  \caption{Colored dots: predictions for the effective mass $|m_{ee}|$
    measured in neutrinoless double beta decay from the neutrino anarchy
    scenario in the type-I seesaw model.  The mass scale of the right-handed
    neutrinos was set to $\mathcal{M} = 10^{14}\ \text{GeV}$ in the left
    panels, and to $\mathcal{M} = 10^{15}\ \text{GeV}$ in the right panels.
    Among the randomly generated parameter points, we distinguish those
    resulting in a normal neutrino mass ordering (upper panels) and those
    resulting in an inverted ordering (lower panels).  Pink points show the
    distributions at $M_\text{GUT}$, while green points show the shifted
    distributions after renormalization group evolution to $M_Z$.  The colored
    bands show the allowed regions in the $m_\text{lightest}$--$|m_{ee}|$ plane
    from the global experimental data on the neutrino mass squared differences
    and mixing angles. Dashed lines correspond to varying only the unknown
    phases in $U_\text{PMNS}$, while solid lines also include the uncertainties
    in the mixing angles.  Note that, to be true to the spirit of anarchy, we
    do not impose any constraints on our randomly generated points, therefore a
    sizeable portion of them fall into regions already excluded by experiment.
    Gray regions are exclusion limits from cosmology~\cite{Ade:2013zuv} and
    from experimental searches for neutrinoless double beta
    decay~\cite{Agostini:2013mzu, Albert:2014awa, Huang:2014qwa}.
    }
  \label{fig:meff1}
\end{figure}

We investigate in \cref{fig:meff1} what neutrino anarchy with and without RG
running predicts for the outcome of future neutrinoless double beta
($0\nu2\beta$) decay experiments.  Of course, since the scale choice
$\mathcal{M}$ for the right-handed neutrinos is an input parameter even in the
anarchy scenario, the absolute scale of $m_\text{lightest}$ and $m_{ee}$
can be chosen at will. In fact, the left and right panels in \cref{fig:meff1}
represent two different choices for $\mathcal{M}$.  Nonetheless, we can draw
several important conclusions from \cref{fig:meff1}.  First, we see again that
anarchy favors the normal mass ordering: there are much fewer points in panels
(c) and (d) than in panels (a) and (b). In other words, if a future
$0\nu2\beta$ search finds a result compatible only with an inverted mass
ordering, this would disfavor anarchy. Second, we see that scenarios with
relatively large $m_{ee}$ but very small $m_\text{lightest}$, such as the
leftmost part of the inverted hierarchy band, are disfavored, as are scenarios
with very small $m_{ee}$ but sizeable $m_\text{lightest}$, in particular
the funnel region in the normal hierarchy case.  Finally, also a
quasi-degenerate mass spectrum is disfavored by the anarchy hypothesis.  If a
combination of $0\nu2\beta$ experiments, cosmology and direct neutrino mass
measurements should point to one of these disfavored regions in the future, the
anarchy hypothesis would come under pressure.

%------------------------------------------------------------------------------
\subsection{Inverse Seesaw Model}
\label{sec:Inverse}      
%------------------------------------------------------------------------------

As a second example, we consider the inverse seesaw scenario~\cite{PhysRevD.34.1642,
PhysRevLett.56.561}, in which the SM Lagrangian is extended by
\begin{align}
  \mathcal{L}_\text{inverse seesaw} \supset
    - (Y_\nu^\dag)^{\alpha\beta} \bar{L}^\alpha \tilde{H} N_R^\beta
    - M_D^{\alpha\beta} \bar{S}^\alpha (N_R^\beta)^c
    - \frac{1}{2} m_s^{\alpha\beta} \bar{S}^\alpha (S^\beta)^c
    + h.c. \,,
  \label{eq:L-inverse-seesaw}
\end{align}
where two types of new gauge singlet Weyl fermions are introduced: $n_N$ generations
of $N_R$ fields, which participate directly in the Yukawa couplings, and $n_S$
generations of $S$ fields. Possible completions of the model that explain this
structure for the Lagrangian are given in \cite{Carvajal:2015dxa, Dias:2012xp,
Dias:2011sq,Dev:2009aw}.
After electroweak symmetry breaking, the neutrino mass term becomes:
\begin{align}
  \mathcal{L}_{m,\text{inverse seesaw}} \supset
   -\big( \overline{\nu_L^c}, \bar N_R, \bar S \big)
    \begin{pmatrix}
      0                  & Y^T_\nu v / \sqrt{2} & 0 \\
      Y_\nu v / \sqrt{2} & 0                    & M^T_D \\
      0                  & M_D                  & m_s
    \end{pmatrix}
    \begin{pmatrix}
      \nu_L \\
      N_R^c \\
      S^c
    \end{pmatrix} \,.
  \label{eq:massISS}
\end{align}
For the effective Majorana mass matrix of the light active neutrinos we
then obtain:
\begin{align}
  m_\nu \simeq \frac{v^2}{2}  Y_\nu^T M_D^{-1} m_s M_D^{-1} Y_\nu \,.
  \label{eq:m-nu-inverse-seesaw}
\end{align}
The correct neutrino mass eigenvalues $\lesssim \mathcal{O}(\text{eV})$ are
obtained for $M_D \sim \mathcal{O}(\text{TeV})$, $m_s \sim
\mathcal{O}(\text{keV})$.  The model has the additional virtue that,
if the number of generations of $S$ is larger than the number
of generations of $N_R$, $\mathcal{O}(\text{keV})$ sterile 
mass eigenstates are obtained \cite{Abada:2014zra}, which could act as the
dark matter in the Universe. (If also $N_R$ has a Majorana mass term, the condition
$n_S > n_N$ can be avoided, see e.g.\ ref.~\cite{Dev:2012bd}.)
The RG equations for the inverse seesaw model are given in \cref{sec:iss-appendix}.
They agree with the ones given in \cite{Bergstrom:2010qb}.

\begin{figure}
  \centering
  \includegraphics{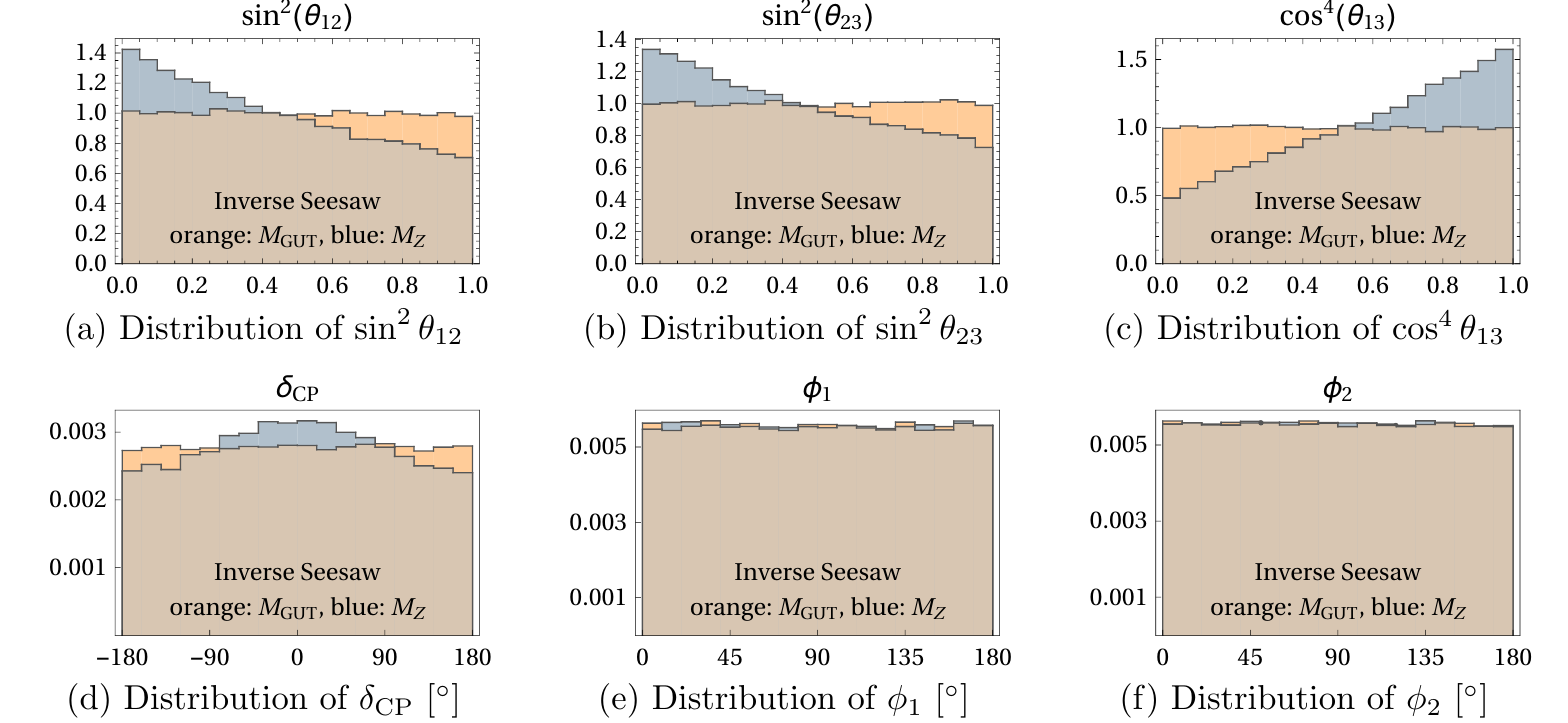}
  \centering
  \caption{Distributions of the mixing angles and physical CP phases
    before and after renormalization group running in the inverse
    seesaw model. The orange regions correspond to the parameters at
    $M_\text{GUT}$, the blue regions to the parameters at $M_Z$.
    As for the type-I seesaw model, the distributions of the Majorana CP
    phases are unaffected by RG evolution. For the mixing angles and the
    Dirac CP phase, the running is much stronger than in the type-I seesaw case.}
  \label{fig:RGEDistISS}
\end{figure}

\begin{figure}
  \centering
  \includegraphics{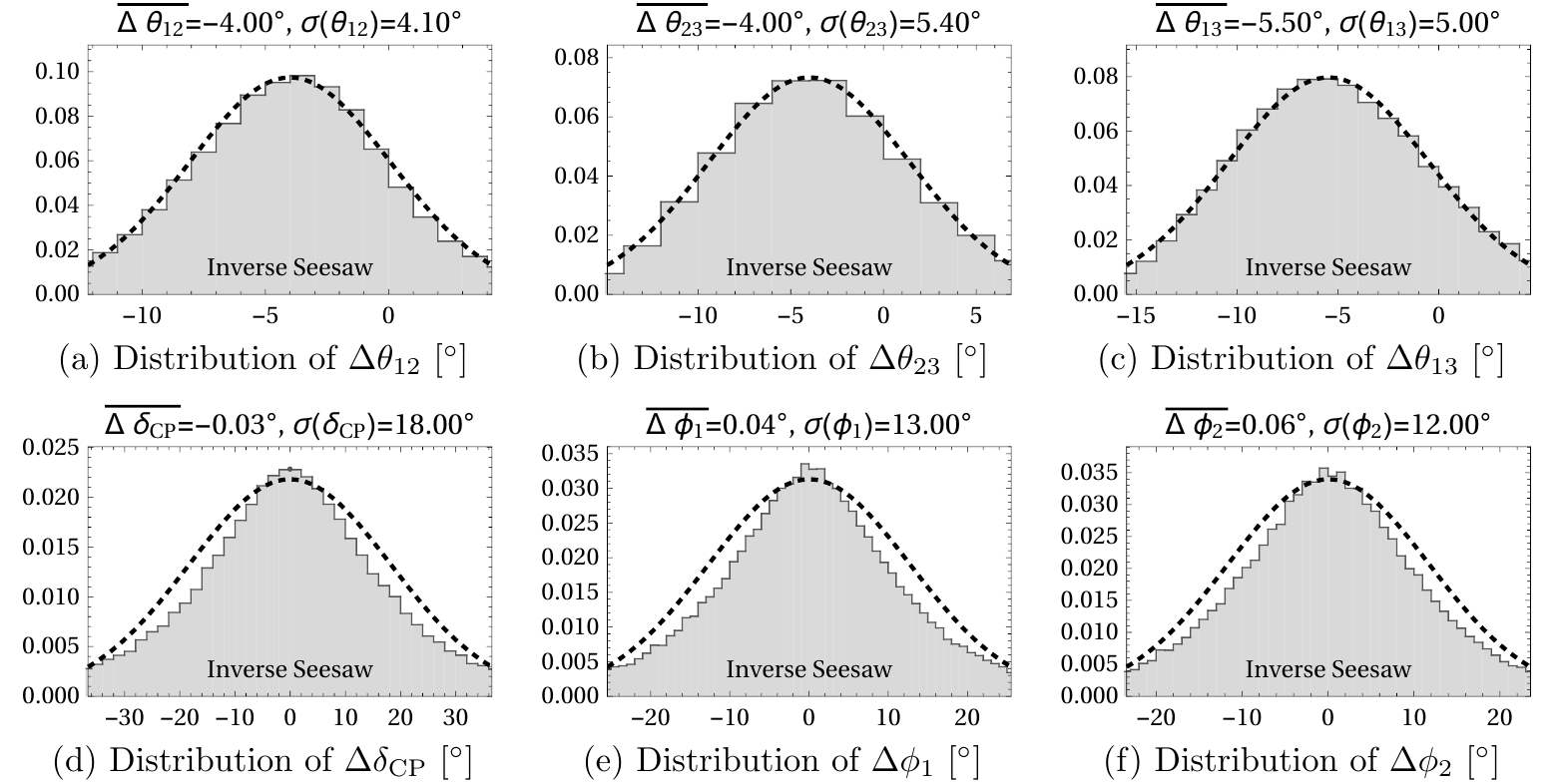}
  \caption{Shift in the mixing angles and the CP phases during RG
    evolution form $M_\text{GUT}$ to $M_Z$ in the inverse seesaw model,
    obtained from $10^5$ randomly generated mass matrices.
    We use the notation $\Delta x \equiv x(M_Z)
    - x(M_\text{GUT})$, where $x$ stands for any of the mixing angles or
    complex phases. Also shown are Gaussian fits with central values
    $\overline{\Delta x}$ and widths $\sigma(\Delta x)$.}
  \label{fig:RGEDeltaISS}
\end{figure}

Carrying out an RG analysis on a random sample of $10^5$
mass matrices of the
form of \cref{eq:massISS}, defined at $M_\text{GUT}$ and evolved down to $M_Z$,
we obtain the results shown in \cref{fig:RGEDistISS,fig:RGEDeltaISS}.  The shifts in the
mixing angles show the same characteristics as in the type-I seesaw model
discussed in the previous section. In particular, mixing angles are preferentially
shifted to smaller values, and $\delta_\text{CP}$ is preferentially close
to zero. The average magnitude of the shifts is, however, larger
than in the type-I seesaw case: of order $4\degree$ for $\theta_{12}$ and
$\theta_{23}$, and of order $5.5\degree$ for $\theta_{13}$.  Also the
widths of the distributions in \cref{fig:RGEDeltaISS} are several times larger
than they were for the type-I seesaw model (see \cref{fig:RGEDeltaSM}).
The average anomalous dimension describing the running of the light
neutrino masses is -0.037 in the inverse seesaw scenario, corresponding to a
70\% decrease of the mass eigenvalues between $M_\text{GUT}$ and $M_Z$.

\begin{figure}
  \centering
  \begin{tabular}{cc}
    \includegraphics[width=0.48\textwidth]{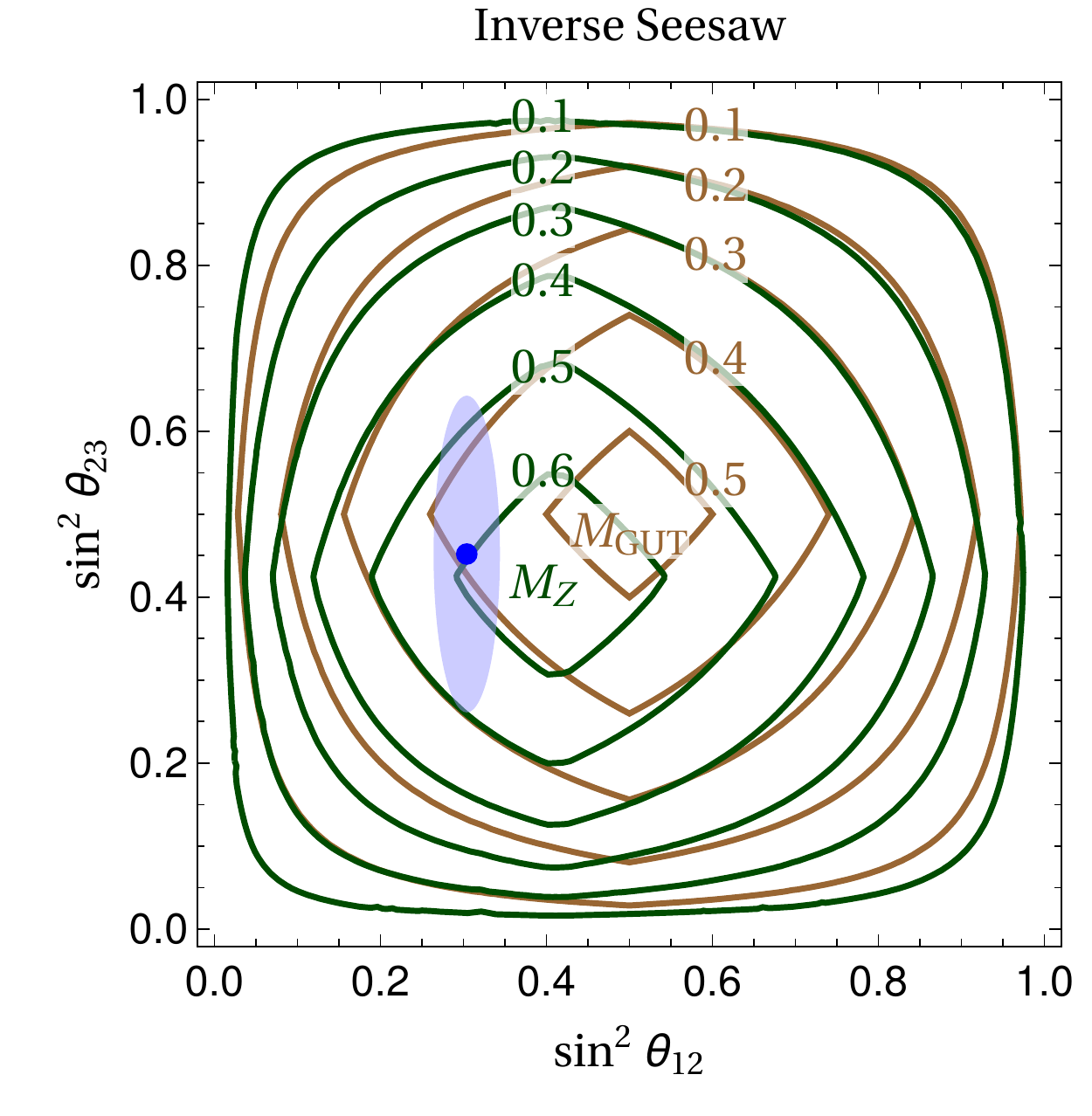} &
    \includegraphics[width=0.48\textwidth]{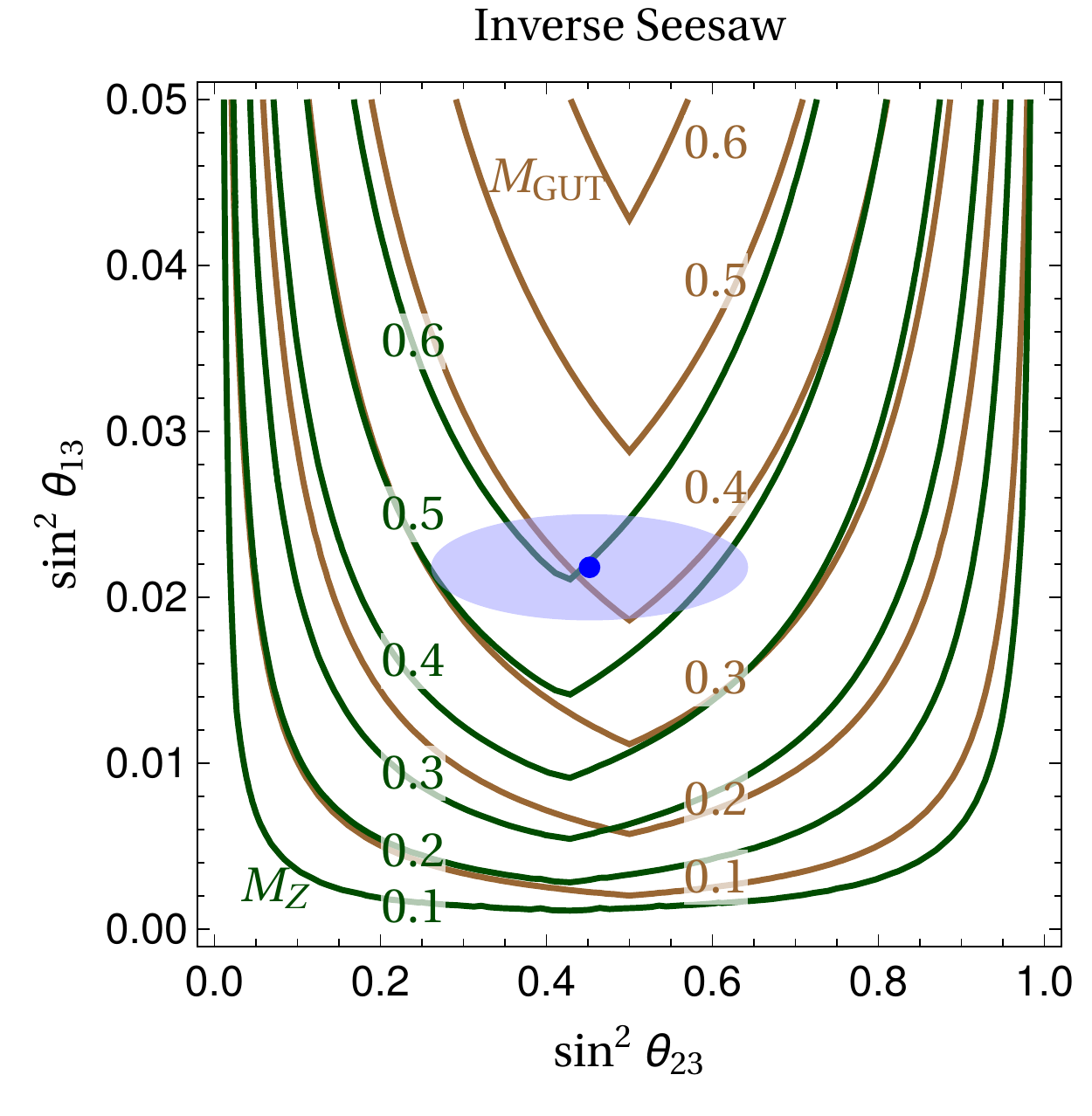} \\
    (a) & (b)
  \end{tabular}
  \caption{(a) Contours of equal probability $P_\text{KS,NH}$ (see
    \cref{eq:P-KS,eq:Ptilde-KS}) in the $\sin^2 \theta_{12}$--$\sin^2
    \theta_{23}$ plane with RG effects included (dark green) and without RG
    effects (brown,see \cref{fig:KS-contours}) for the inverse seesaw model. We
    consider only parameter points with a normal mass ordering (see text in
    \cref{sec:TypeI} for details).  The blue ellipses denote the favored
    parameter region from the global fit in
    ref.~\citep{Gonzalez-Garcia:2014bfa} at the $3\sigma$ confidence level, and
    the blue dot is the corresponding best fit point.  We
    have fixed the third mixing angle at its best fit value, $\sin^2
    \theta_{13} = 0.0218$.  With RG effects included, the likelihood of the
    data with respect to the anarchy hypothesis increases from 41.1\% to 59.3\%.
    (b) Contours of equal probability $P_\text{KS,NH}$ in the $\sin^2
    \theta_{23}$--$\sin^2 \theta_{13}$ plane with and without RG effects
    included (same coloring as in (a)).  The solar mixing angle is fixed at its
    best fit value, $\sin^2 \theta_{12} = 0.304$.}
  \label{fig:rgeVersusNoRGEIS}
\end{figure}

To see how RG effects influence the consistency of the neutrino oscillation
data with the predictions of anarchy, we compare in \cref{fig:rgeVersusNoRGEIS}
the equiprobability contours with and without RG running in the inverse
seesaw model. To produce this plot, we follow the same procedure as for the
type-I seesaw model (see \cref{sec:TypeI}). We exploit again the fact that
the mixing parameters are still nearly uncorrelated after RG evolution, even
though the absolute values of the correlation coefficients are now somewhat
larger, $\sim 0.07$ now.  We find that RG effects change the compatibility of the
anarchy hypothesis with the data substantially: $\tilde{P}_\text{KS}$ increases from
41.1\% without RG effects to 59.3\% with RG running included.

%------------------------------------------------------------------------------
\subsection{\texorpdfstring{Inverse Seesaw model with gauged $\boldsymbol{U(1)_{B-L}}$}
           {Inverse Seesaw model with gauged B-L}}
\label{sec:ISSBL}
%------------------------------------------------------------------------------

\begin{table}
  \renewcommand{\arraystretch}{1.3}
  \begin{minipage}{8cm}
    \begin{ruledtabular}
    \begin{tabular}{ccc}
      field      &      Spin     & $U(1)_{B-L}$ charge $Q_{B-L}$ \\ \hline
      $\chi$     &       0       & $-1$                          \\
      $N_R$      & $\frac{1}{2}$ & $-1$                          \\
      $S$        & $\frac{1}{2}$ & $+2$                          \\
      $S'$       & $\frac{1}{2}$ & $-2$                          \\
      $Z_{\mu}'$ &       1       & $0$
    \end{tabular}
    \end{ruledtabular}
  \end{minipage}
  \caption{The new field content of the inverse seesaw model with gauged
    $U(1)_{B-L}$. All listed fields are SM singlets. As usual, standard model leptons
    have $B-L$ charge $-1$, while quarks have $B-L$ charges of $1/3$.}
  \label{tab:B-L-fields}
\end{table}

As an example for a specific realization of the inverse seesaw mechanism, we
consider an extension of the SM where the difference of baryon number and
lepton number, $B-L$, is gauged~\cite{Khalil:2010iu}. The corresponding gauge
boson $Z'$ is assumed to have a mass of order $100\ \text{TeV}$ generated by
the vacuum expectation value (vev) of a new scalar $\chi$ with quantum numbers
$(1, 1, 0, -1)$ under $SU(3)_c \times SU(2)_L \times U(1)_Y \times U(1)_{B-L}$.
The fermionic field content of the model is the same as in the phenomenological
inverse seesaw model discussed in \cref{sec:Inverse}
($n_N$ singlets $N_R$, $n_S$ singlets $S$), but the fermions now carry $B-L$
charge: the $N_R$ have $B-L$ charge $Q_{B-L} = -1$, while
the $S$ have $Q_{B-L} = +2$.  To ensure
anomaly cancellation, it is necessary to supplement the model by $n_S$ extra
Weyl fermions $S'$ with $B-L$ charge $-2$.\footnote{The cancellation of the
  triangular gauge anomaly requires that $\sum_{\psi_{L}} (Q^{\psi_L}_{B-L})^3
  = \sum_{\psi_{R}} (Q^{\psi_{R}}_{B-L})^3$, where $\psi_{L}$ and $\psi_{R}$
represent left-handed and right-handed fields, respectively.}
For simplicity, we assume that couplings of $S'$
to other fields (except the $U(1)_{B-L}$ gauge boson) are forbidden by a new
discrete symmetry.  We will here remain agnostic with regard to the details of
this symmetry and simply ignore $S'$ in our subsequent discussion.  The quantum
numbers of the new particles in the inverse seesaw model with gauged $B-L$ are
summarized in \cref{tab:B-L-fields}.

The Lagrangian reads \cite{Khalil:2010iu}
\begin{multline}
  \mathcal{L} \supset
              - \frac{1}{4} Z'_{\mu\nu} Z'^{\mu\nu}
              + \!\! \sum_{\psi=N_R,S,S'} \!\!\! i\,\bar\psi \slashed{D} \psi
              + (D_{\mu} \chi) (D^{\mu} \chi)^{\dagger}
              - \big[ (Y_\nu^\dag)^{\alpha\beta} \bar{L}_L^\alpha \tilde{H} \, N_R^\beta
                    + \lambda_S^{\alpha\beta} \, \bar{S}^\alpha \chi^\dagger (N_R^\beta)^c
                    + h.c. \big] \\
              + \mu_2 \, \chi^\dagger \chi
              + \lambda_2 \, (\chi^\dagger \chi)(H^\dagger H)
              + \lambda_3 \ (\chi^{\dagger} \chi)^2
              + \frac{1}{\Lambda^3} \overline{S^c} \chi^4 S \,,
  \label{eq:LagBL}
\end{multline}
where $D_\mu = \partial_\mu - i g_{B-L} Q_{B-L} Z^\prime_{\mu}$ is
the covariant derivative.
Note that the last term in the second row is dimension 7 and is therefore
suppressed by three powers of a cutoff scale $\Lambda$.\footnote{A realization
of this term in a renormalizable model requires additional fields.
As a proof of principle, we could add a Majorana fermion $N_R'$ and a real scalar $\rho$,
both of which are total singlets under the SM gauge group and under $U(1)_{B-L}$,
plus a complex scalar $\eta$ with $B-L$ charge $-2$. If $N_R'$, $\rho$ and
$\eta$ are charged under a new $\mathbb{Z}_2$ symmetry under
which all other fields are even, the operator $\overline{S^c} \chi^4 S / \Lambda^3$
is generated through the following diagram:%
\begin{center}
  \includegraphics[scale=0.7]{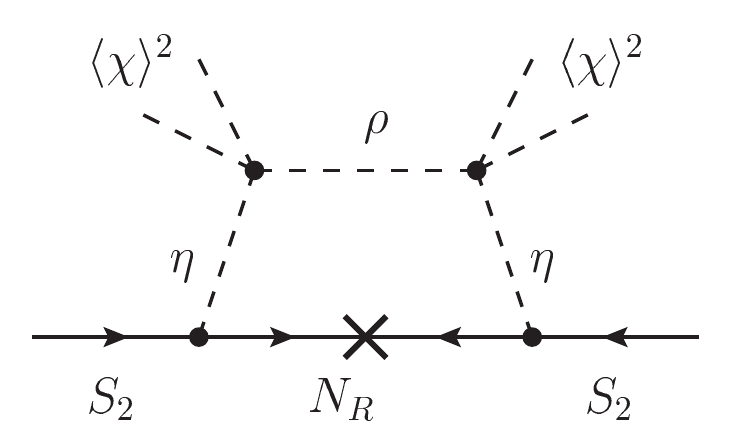}
\end{center}
}
We have not written down other higher-dimensional operators here since they
are not relevant to the phenomenology we are interested in.

As the vev of the new scalar $\chi$ breaks the $U(1)_{B-L}$ symmetry
spontaneously, it endows the $B-L$ gauge boson with a mass $M_{Z^\prime} =
\sqrt{2}g_{B-L} \ev{\chi}$.  Moreover, it generates a Dirac-type mass mixing term of
the form $(M_D)^{\alpha\beta} \bar{S}^\alpha (N_R^\beta)^c$ with $M_D =
\lambda_S \ev{\chi^\dagger}$, and a Majorana mass term $m_s \overline{S^c} S$
with $m_s = \ev{\chi}^4 / \Lambda^3$. Note that, in the unbroken phase of $U(1)_{B-L}$
a Majorana mass term for $S$ is forbidden due to its nonzero $B-L$ charge.
After $U(1)_{B-L}$-breaking and
electroweak symmetry breaking, the neutral components of the SM Higgs doublet
$H$ and the new scalar $\chi$ mix to form two $CP$ even
scalars~\cite{Emam:2007dy}.

\begin{figure}
  \centering
  \includegraphics[width=0.6\textwidth]{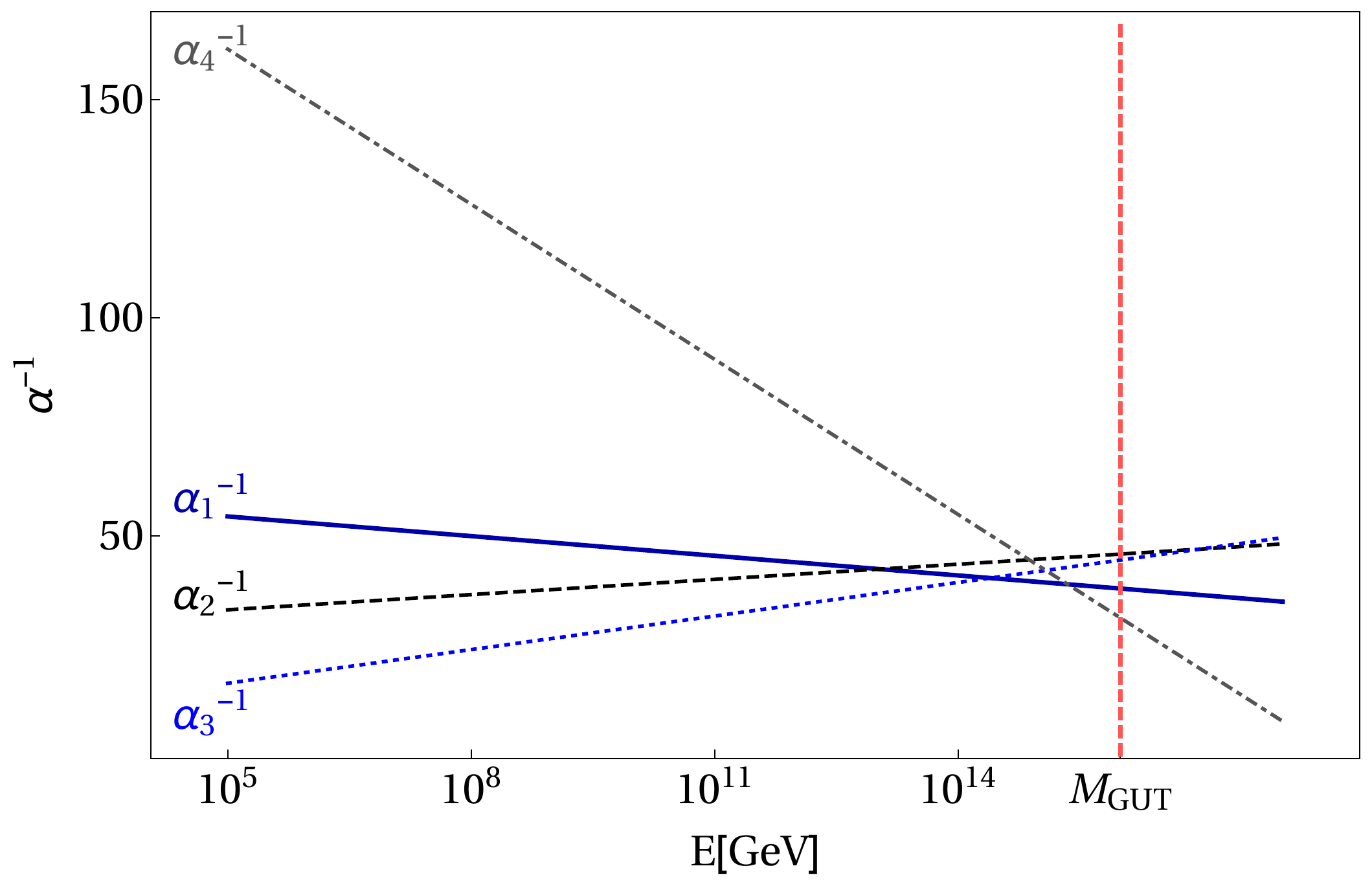}
  \caption{Running of the fine structure constant $\alpha_1$ for $U(1)_Y$
    (dark blue solid), $\alpha_2$ for $SU(2)_L$ (black dashed), $\alpha_3$ for
    $SU(3)_c$ (blue dotted), and $\alpha_4 \equiv g_{B-L}^2 / (4\pi)$ for
    $U(1)_{B-L}$ (dark gray dot-dashed) in the inverse seesaw model with gauged
    $B-L$.}
  \label{fig:running-BL}
\end{figure}

Since our main goal in this section is only to explore whether a UV completion
of the inverse seesaw model leads to qualitatively different RG effects, we
make several simplifying assumptions in our RG analysis.
First, to avoid having to deal with the
multitude of different effective theories that can arise depending on where
$B-L$ is broken relative to the masses of the heavy $\mathcal{O}(\text{TeV})$
neutrinos, we always fix the $U(1)_{B-L}$ breaking scale at $f_{B-L} =
100$~TeV, well above the heavy neutrino masses. Moreover, we assume for
simplicity that the $B-L$ breaking scale $f_{B-L}$, the vev $\ev{\chi}$, and
the mass $m_\chi$ of $\chi$ are identical, so that we do not need to consider
running between these scales.  We choose the numerical value of $g_{B-L}$ such
that it unifies to the extent possible at
the GUT scale with the SM gauge couplings, see \cref{fig:running-BL}.
We find that a reasonable choice is $g_{B-L}=0.58$ at the GUT scale.
By demanding that the running $Z'$ mass parameter $M_{Z'}(\mu) \equiv
\sqrt{2} g_{B-L}(\mu) \ev{\chi}$ equals the scale $\mu$, we find the
threshold mass $M_{Z^\prime} \approx 39~\text{TeV}$.  
Since this is sufficiently close to $f_{B-L} = 100~\TeV$ on a log
scale, we integrate out $\chi$ and $Z'$ simultaneously.  As a further
simplification, we ignore renormalization group effects on the quadratic terms
in the scalar potential, and consequently the scale dependence of the scalar
vevs. Since the scalar quartic couplings do not affect the flavor structure,
their only influence on the neutrino mixing parameter can be indirect,
by changing the various mass scales of the theory.
Finally, we neglect kinetic mixing between the $Z^\prime$ and the standard
model hypercharge gauge boson $B$, which would be allowed by the
symmetries.  Kinetic mixing is in general absent at the grand unification
scale, where we assume that $U(1)_{B-L}$ emerges from a larger, non-Abelian
gauge group. Even though kinetic mixing does get induced in the renormalization
group evolution to lower scales at one loop level, its effect on other running
parameters is suppressed by two loop factors and therefore negligible compared
to other parameters appearing in the RG equations.  In the same spirit
we will also assume the Higgs portal coupling $\lambda_2$ to be strongly suppressed.

With these simplifications, the effective theory valid at scales $\mu < f_{B-L}$
is simply the one discussed in \cref{sec:Inverse}. We thus only need to evolve
the parameters of the full model from $M_\text{GUT}$ to $f_{B-L}$ and then feed
them into the previously discussed evolution equations for the phenomenological
inverse seesaw model from \cref{sec:Inverse}.
The renormalization group equations for the inverse seesaw model with gauged
$B-L$ are given in \cref{sec:iss-B-L}. While their form is independent
of the number of generations of singlet fields we introduce, we choose to
invoke three $N_R$ fields and four $S$ fields, yielding one $\mathcal{O}(\text{keV})$
mass-eigenstate as a possible DM candidate.

\begin{figure}
  \centering
  \includegraphics{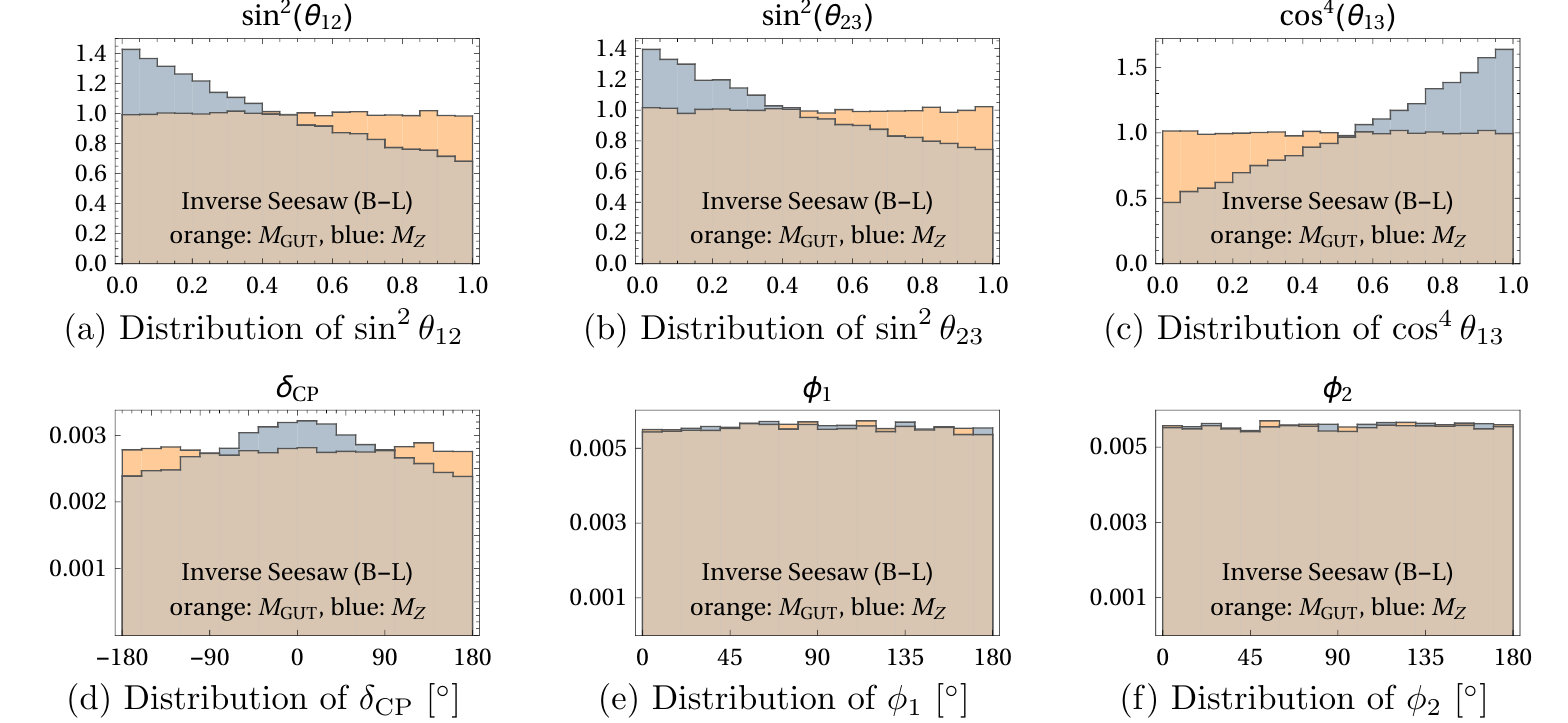}
  \centering
  \caption{Distributions of the mixing angles and physical CP phases
    before and after renormalization group running in the inverse seesaw model
    with gauged $B-L$. The orange regions correspond to the parameters at
    $M_\text{GUT}$,h scale, the blue regions to the parameters at $M_Z$.  We
    find that the distributions are practically identical to those in the
    phenomenological inverse seesaw model from \cref{sec:Inverse}, see
    \cref{fig:RGEDistISS}.}
  \label{fig:RGEDistBL}
\end{figure}

\begin{figure}
  \centering
  \includegraphics{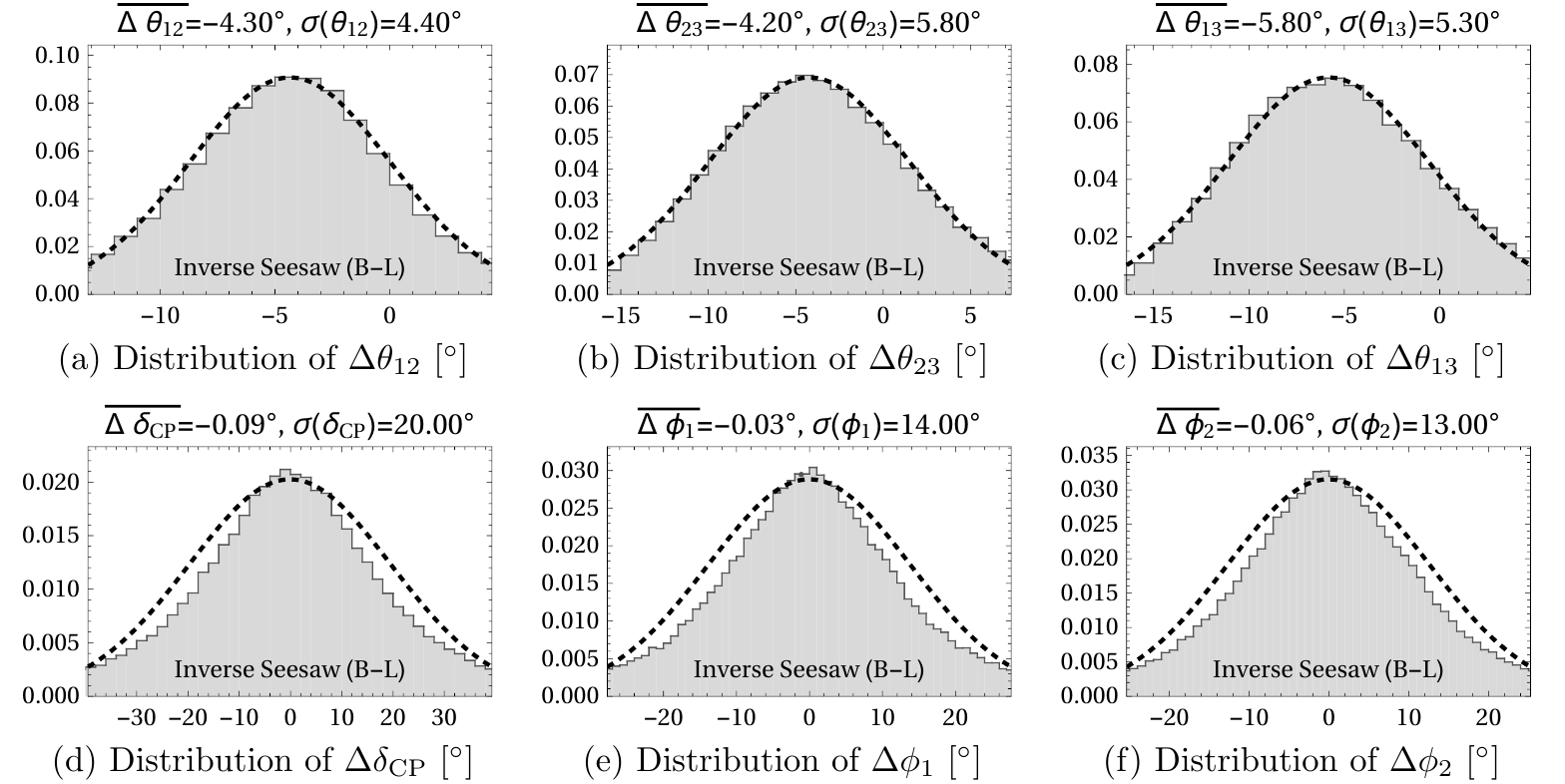}
  \caption{Shifts in the mixing angles and the CP phases during RG evolution
    form $M_\text{GUT}$ to $M_Z$ in the inverse seesaw model with gauged $B-L$,
    obtained from $10^5$ randomly generated mass matrices. We use the notation
    $\Delta x \equiv x(M_Z) - x(M_\text{GUT})$, where $x$ stands for any of the
    mixing angles or complex phases. Also shown are Gaussian fits with central
    values $\overline{\Delta x}$ and widths $\sigma(\Delta x)$.
    The distributions are practically identical to those in
    the phenomenological inverse seesaw model from \cref{sec:Inverse},
    see \cref{fig:RGEDeltaISS}.}
  \label{fig:RGEDeltaBL}
\end{figure}

Proceeding as we did in \cref{sec:TypeI,sec:Inverse}, we investigate how the
statistical distributions of the neutrino mixing parameters change due to RG
effects in the inverse seesaw model with gauged $B-L$.
We generate $10^5$ parameter sets at $M_\text{GUT}$ and evolve them down
to $M_Z$. In \cref{fig:RGEDistBL} we show the distributions of the mixing parameters
before and after running, and in \cref{fig:RGEDeltaBL} we plot the distributions
of the RG-induced shifts $\Delta x$.
We find that these distributions are practically identical
to those in the phenomenological inverse seesaw model without gauged \text{$B-L$}
number. Consequently, also the value of $\tilde{P}_\text{KS}$ describing
the compatibility of the data with anarchy is similar: 60.4\%, compared to
59.3\% in the phenomenological model.
The running of the light neutrino masses is somewhat larger than in the
phenomenological model, though: the average anomalous dimensions is -0.051, corresponding
to a decrease of the mass eigenvalues by 80\%.

%==============================================================================
\section{Summary and Conclusions}
\label{sec:conclusions}
%==============================================================================

In summary, we have studied how renormalization group effects alter the
predictions of neutrino anarchy. We have found that the mixing angles at $M_Z$
are on average several degrees smaller than at $M_\text{GUT}$, and that the
statistical distribution of the Dirac CP phase $\delta_\text{CP}$ after RG
running peaks at zero. The Kolmogorov--Smirnov $p$-value describing the
compatibility of the observed mixing angles with the anarchy hypothesis thus
increases from 41.1\% without RG running to 47.7\% with RG running in the type-I
seesaw model, to 59.3\% with RG running in the phenomenological inverse seesaw model,
and to 60.4\% with RG running in the inverse seesaw model with gauged $B-L$.
Our results thus strengthen the claim that anarchy offers a good
description of neutrino oscillation data, and they demonstrate that RG effects
are crucial and need to be taken into account when deriving quantitative
predictions in the anarchy scenario.

%==============================================================================
\section*{Acknowledgments}
%==============================================================================

It is a pleasure to thank Andr\'e de Gouvea, J\"orn Kersten, and Hitoshi
Murayama for useful discussions. We are moreover grateful to Alexander Merle for
help in generating \cref{fig:meff1}, based on his papers \cite{Lindner:2005kr,
Merle:2006du}.  The work of VB and JK is supported by the German Research
Foundation (DFG) under Grant Nos.\ \mbox{KO~4820/1--1} and FOR~2239 and by the
European Research Council (ERC) under the European Union's Horizon 2020
research and innovation programme (grant agreement No.\ 637506,
``$\nu$Directions'').  VB and MK are supported by the DFG Graduate School
Symmetry Breaking in Fundamental Interactions (GRK 1581).  Additional support
has been provided by the Cluster of Excellence ``Precision Physics, Fundamental
Interactions and Structure of Matter'' (PRISMA -- EXC 1098), grant No.~05H12UME
of the German Federal Ministry for Education and Research (BMBF).

%==============================================================================
\appendix
\section{Renormalization group equations for the type-I seesaw
         and inverse seesaw models}
\label{sec:iss-appendix}
%==============================================================================

In this appendix, we list the renormalization group equations for the
inverse seesaw model discussed in \cref{sec:Inverse}.  By simply setting
certain terms to zero in these equations, also the RG equations for
the type-I seesaw will follow (see \cref{sec:type-I-from-iss}).

%------------------------------------------------------------------------------
\subsection{Structure of the mass matrix}
\label{sec:iss-mass-matrix}
%------------------------------------------------------------------------------

We start from the mass matrix of the inverse seesaw model,
eq.~\eqref{eq:massISS}, which we repeat here:
\begin{align}
  M_\nu \equiv
    \begin{pmatrix}
      0                  & Y^T_\nu v / \sqrt{2} & 0 \\
      Y_\nu v / \sqrt{2} & 0                    & M^T_D \\
      0                  & M_D                  & m_s
    \end{pmatrix} \,.
  \label{eq:M-nu-iss}
\end{align}
For 3 generations of active neutrinos, $n_N$ generations of heavy $N_R$ fields,
and $n_S$ generations of $S$ fields, $Y_\nu$ is an $n_N \times 3$ matrix,
$M_D$ is an $n_S \times n_N$ matrix, and $m_s$ is an $n_S \times n_S$ matrix.
We assume $\| Y_\nu \| \sim \mathcal{O}(1)$, $\| M_D \| \sim \text{TeV}$
and $\| m_s \| \sim \text{keV}$. We moreover assume $n_S > n_N$ so that
$M_\nu$ has at least one $\mathcal{O}(\text{keV})$ eigenvalue, providing a
dark matter candidate.

For the RG analysis, it is more convenient to work with a matrix which has
diagonal elements much larger than the off-diagonal ones. Moreover, it is
convenient to group the elements of this matrix by their typical scale. We
therefore apply a unitary transformation $V$ to $M_\nu$ to bring it to the
form
\begin{align}
  \tilde{M}_\nu \equiv
    V^T M_\nu V =
    \begin{pmatrix}
      0                          & \hat{Y}^T_\nu v / \sqrt{2} & \tilde{Y}^T_\nu v / \sqrt{2} \\
      \hat{Y}_\nu v / \sqrt{2}   & M_N                        & m^T_D \\
      \tilde{Y}_\nu v / \sqrt{2} & m_D                        & \hat{m}_s
    \end{pmatrix} \,.
  \label{eq:M-nu-tilde}
\end{align}
$V$ is chosen such that $\| M_N \| \gg \|m_D\|$, $\| \hat{m}_s \sim \text{keV} \|$
and $\| \hat{Y}_\nu \| \sim \| Y_\nu \| \gg \| \tilde{Y}_\nu \|$.
Thus, $M_N$ is a $2 n_N \times 2
n_N$ matrix with eigenvalues of order TeV and $\hat{m}_s$ is an $(n_S - n_N) \times
(n_S - n_N)$ matrix.

To construct $V$ explicitly, we first consider the unitary $n_S \times n_S$ matrix
$u_L$ and the unitary $n_N \times n_N$ matrix $u_R$ which diagonalize
$M_D$ according to
\begin{align}
  M_D^\text{diag} = u_L^\dag M_D u_R \,.
\end{align}
Here $M_D^\text{diag}$ means the diagonal matrix of eigenvalues, \emph{ordered
in descending order}.  From $u_L$ and $u_R$, we construct
\begin{align}
  U_L &= \begin{pmatrix}
           \mathbf{1}_{3+n_N} & 0 \\
           0                  & u_L
         \end{pmatrix}
  &
  U_R &= \begin{pmatrix}
           \mathbf{1}_{3} & 0   & 0 \\
           0              & u_R & 0 \\
           0              & 0   & \mathbf{1}_{n_S}
         \end{pmatrix}
  &
  W &= U_R U_L^* \,.
\end{align}
Then, the transformation
\begin{align}
  W^T M_\nu W
\end{align}
will diagonalize the $M_D$ blocks in $m_\nu$, putting the largest mass eigenvalues
on top.  After this, we perform a rotation by $\pi/4$ in the singlet block, given by:
\begin{align}
  X &= \begin{pmatrix}
         \mathbf{1}_3 & 0                   & 0 \\
         0            &  r_{n_N \times n_N} & r_{n_N \times n_S} \\
         0            & -r_{n_S \times n_N} & r_{n_S \times n_S}
       \end{pmatrix} \,,
  \label{eq:ISSrotX}
\end{align}
where
\begin{align}
  r_{n \times m} \equiv \frac{1}{\sqrt{2}}
    \begin{pmatrix}
      1      & \cdots & 0      & 0      & \cdots & 0      \\
      \vdots & \ddots & \vdots & \vdots & \ddots & \vdots \\
      0      & \cdots & 1      & 0      & \cdots & 0
    \end{pmatrix}
\end{align}
is the $n \times n$ identity matrix (multiplied by $1/\sqrt{2}$), padded with zeroes to 
turn it into an $n \times m$ matrix.
After applying the full rotation
\begin{align}
 V \equiv U_R U_L^* X
\end{align}
to $M_\nu$, we obtain $\tilde{M}_\nu$ in the form given in
eq.~\eqref{eq:M-nu-tilde}.  We summarize the dimensions and the mass scales for
the different blocks of $M_\nu$ and $\tilde{M}_\nu$ in \cref{tab:dimMass}.  The
$2n_N \times 2n_N$ matrix $M_N$ has $n_N$ pairs of eigenvalues that differ only
by $\Delta M \sim \mathcal{O}(\text{keV})$, much less than the absolute scale
of the eigenvalues of order TeV. The eigenstates in each pair couple to the
active neutrinos via Yukawa couplings that are equal (up to a factor of $i$).
Their contributions to the light neutrino masses therefore cancel up to terms of
order $\Delta M / \|M_N\|$.  This explains the smallness of the active neutrino
masses despite the fact that $\hat{Y}_\nu$ is $\mathcal{O}(1)$ and $M_N$ is $\mathcal
O(\mathrm{TeV})$.  Another small contribution to the active neutrino masses comes
from the $\mathcal{O}(\text{keV})$ mass eigenstate mixing with the active
fields through the strongly suppressed $\tilde Y_\nu$.  We can also understand
why $M_\nu$ has $n_S - n_N$ eigenvalues of order keV: the off-diagonal elements
in the last $n_S-n_N$ rows and columns of $X$ are zero and thus we get
$n_S-n_N$ keV states that do not mix with the $N_R$ or $\nu_L$ fields.

\begin{table}
  \setlength{\tabcolsep}{0.5cm}
  \renewcommand{\arraystretch}{1.5}
  \newcommand{\mcx}[3]{\multicolumn{#1}{#2|}{#3}}
  \newcommand{\mc}[3]{\multicolumn{#1}{#2}{#3}}
  \begin{center}
    \begin{ruledtabular}
    \begin{tabular}{cc|ccc}
      \mcx{2}{c}{Original basis} & \mc{3}{c}{Rotated basis} \\
      \hline
      matrix & dimensions & matrix & dimensions & scale    \\
      $Y_\nu$ & $n_N \times 3$   & $\hat{Y}_\nu$  & $2n_N \times 3$              & $\mathcal{O}(Y_\nu)$         \\
      $M_D$   & $n_S \times n_N$ & $\tilde Y_\nu$ & $(n_S-n_N) \times 3$         & $\mathcal{O}(Y_\nu m_s/M_D)$ \\
      $m_s$ & $n_S \times n_S$   & $M_N$          & $2n_N \times 2n_N$           & $\mathcal{O}(M_D)$           \\
              &                  & $m_D$          & $(n_S-n_N) \times 2n_N$      & $\mathcal{O}(m_s)$           \\
              &                  & $\hat{m}_s$    & $(n_S-n_N) \times (n_S-n_N)$ & $\mathcal{O}(m_s)$
    \end{tabular}
  \end{ruledtabular}
  \end{center}
  \caption{Dimensions of the submatrices forming the mass matrices $M_\nu$
    and $\tilde{M}_\nu$ in the inverse seesaw model (see
    \cref{eq:M-nu-iss,eq:M-nu-tilde}).  In the last column, we indicate the typical
    mass scale of the matrices in the rotated basis, \cref{eq:M-nu-tilde}.}
  \label{tab:dimMass}
\end{table}

%------------------------------------------------------------------------------
\subsection{Beta functions for the inverse seesaw model}
\label{sec:iss-beta-functions}
%------------------------------------------------------------------------------

We derive the RG equation for the mass matrix $\tilde{M}_\nu$
given in \cref{eq:M-nu-tilde} by using the fact that $\tilde{M}_\nu$
has exactly the same structure as the mass matrix of a type-I seesaw model.

Whenever the renormalization scale $\mu$ crosses the mass scale of one of the
heavy sterile neutrinos as it is evolved from $M_\text{GUT}$ to $M_Z$, the
corresponding row and column is removed from the mass matrix $\tilde{M}_\nu$.
Moreover the remaining rows and columns are modified to include the effect of
integrating out one neutrino state. This in particular implies that the upper
$3 \times 3$ block of $\tilde{M}_\nu$, which is zero initially, becomes a
nonzero matrix $\kappa$. (We follow the notation of ref.~\cite{Antusch:2005gp}
here.) If we did run down to sub-keV scales
and integrated out all sterile states, $\kappa$ would eventually coincide with
$m_\nu = \frac{v^2}{2} Y_\nu^T M_D^{-1} m_s M_D^{-1} Y_\nu$.

For the $\beta$ functions of a parameter $x$, we use the convention
\begin{align}
  \beta_x = \mu \frac{d}{d\mu} x \,.
\end{align}
The evolution equations read
\begin{align}
  16 \pi^2 \beta_{M_N} &=
    \big( \hat{Y}_\nu \hat{Y}_\nu^\dagger \big) M_N
    + M_N \big(\hat{Y}_\nu \hat{Y}_\nu^\dagger \big)^T
    + \big( \hat{Y}_\nu \tilde{Y}_\nu^\dagger \big) m_D
    + m_D^T \big(\hat{Y}_\nu \tilde{Y}_\nu^\dagger \big)^T
  \label{eq:ISSbetafirst} \\
  16 \pi^2 \beta_{\hat{m}_s} &=
      \big( \tilde{Y}_\nu \tilde{Y}_\nu^\dagger \big)\hat{m}_s
    + \hat{m}_s \big( \tilde{Y}_\nu \tilde{Y}_\nu^\dagger \big)^T
    + \big( \tilde{Y}_\nu \hat{Y}_\nu^\dagger \big) m_D^T
    + m_D \big( \tilde{Y}_\nu \hat{Y}_\nu^\dagger \big)^T \\
  16 \pi^2 \beta_{m_D} &=
      \big( \tilde{Y}_\nu \tilde{Y}_\nu^\dagger \big) m_D
    + m_D \big( \hat{Y}_\nu \hat{Y}_\nu^\dagger \big)^T
    + \big(\tilde{Y}_\nu \hat{Y}_\nu^\dagger \big) M
    + \hat{m}_s \big( \hat{Y}_\nu \tilde{Y}_\nu^\dagger \big)^T \\
  16 \pi^2 \beta_{\kappa} &=
    - \frac{3}{2} \big( Y_e^\dagger Y_e \big)^T \kappa
    - \frac{3}{2} \kappa \big( Y_e^\dagger Y_e \big)
    + \frac{1}{2} \big( \hat{Y}_\nu^\dagger \hat{Y}_\nu \big)^T \kappa
    + \frac{1}{2}\kappa \big( \hat{Y}_\nu^\dagger \hat{Y}_\nu \big) \nonumber\\
   &\quad+ 2\,\tr \big( Y_e^\dagger Y_e \big) \kappa
    + 2\,\tr \big( Y_\nu^\dagger Y_\nu \big) \kappa
    + 6\,\tr \big( Y_u^\dagger Y_u \big) \kappa \nonumber\\
   &\quad+ 6\,\tr \big( Y_d^\dagger Y_d \big) \kappa
    - 3 g_2^2 \kappa
    + \lambda\kappa \\
  16 \pi^2 \beta_{\hat{Y}_\nu} &= \hat{Y}_\nu
    \bigg[  \frac{3}{2} \big( \tilde{Y}_\nu^\dagger \tilde{Y}_\nu
                            + \hat{Y}_\nu^\dagger \hat{Y}_\nu \big)
          - \frac{3}{2} \big( Y_e^\dagger Y_e \big)
          + \tr\big( \hat{Y}_\nu^\dagger \hat{Y}_\nu
                   + \tilde Y_\nu^\dagger \tilde Y_\nu \big) \nonumber\\
 &\qquad\ + \tr\big( Y_e^\dagger Y_e \big)
          + 3\,\tr\big( Y_u^\dagger Y_u \big)
          + 3\,\tr\big( Y_d^\dagger Y_d \big)
          - \frac{9}{20} g_1^2
          - \frac{9}{4} g_2^2 \bigg] \\
  16 \pi^2 \beta_{\tilde Y_\nu} &= \tilde Y_\nu
    \bigg[ \frac{3}{2} \big( \tilde Y_\nu^\dagger \tilde Y_\nu
                           + \hat{Y}_\nu^\dagger \hat{Y}_\nu \big)
         - \frac{3}{2} \big( Y_e^\dagger Y_e \big)
         + \tr\big( \hat{Y}_\nu^\dagger \hat{Y}_\nu
         + \tilde Y_\nu^\dagger \tilde Y_\nu \big)  \nonumber \\
  &\qquad\ + \tr\big( Y_e^\dagger Y_e \big)
         + 3\,\tr\big( Y_u^\dagger Y_u \big)
         + 3\,\tr\big(Y _d^\dagger Y_d \big)
         - \frac{9}{20} g_1^2
         - \frac{9}{4}g_2^2 \bigg] \\
  16 \pi^2 \beta_{Y_e} &= Y_{e}
    \bigg[ \frac{3}{2} Y_e^\dagger Y_e
         - \frac{3}{2} \big( \hat{Y}_\nu^\dagger \hat{Y}_\nu
                           + \tilde Y_\nu^\dagger \tilde Y_\nu \big)
         - \frac{9}{4} g_1^2
         - \frac{9}{4} g_2^2 \nonumber\\ 
  &\qquad\ + \tr\big[ Y_e^\dagger Y_e
                    + \hat{Y}_\nu^\dagger \hat{Y}_\nu
                    + \tilde Y_\nu^\dagger \tilde Y_\nu
                    + 3 Y_d^\dagger Y_d
                    + 3 Y_u^\dagger Y_u \big] \bigg] \\
  16 \pi^2 \beta_{Y_d} &= Y_d
    \bigg[ \frac{3}{2} Y_d^\dagger Y_d
         - \frac{3}{2} Y_u^\dagger Y_u
         - \frac{1}{4} g_1^2
         - \frac{9}{4} g_2^2
         - 8g_3^2   \nonumber\\
  &\qquad\ + \tr\big[ Y_e^\dagger Y_e
                    + \hat{Y}_\nu^\dagger \hat{Y}_\nu
                    + \tilde Y_\nu ^\dagger \tilde Y_\nu
                    + 3 Y_d^\dagger Y_d
                    + 3 Y_u^\dagger Y_u \big] \bigg] \\
  16 \pi^2 \beta_{Y_u} &= Y_u
    \bigg[ \frac{3}{2} Y_u^\dagger Y_u
         - \frac{3}{2} Y_d^\dagger Y_d
         - \frac{17}{20} g_1^2
         - \frac{9}{4} g_2^2
         - 8 g_3^2   \nonumber\\
  &\qquad\ + \tr\big[ Y_e^\dagger Y_e
                    + \hat{Y}_\nu^\dagger \hat{Y}_\nu
                    + \tilde Y_\nu^\dagger \tilde Y_\nu
                    + 3 Y_d^\dagger Y_d
                    + 3 Y_u^\dagger Y_u \big] \bigg] \\
  16 \pi^2 \beta_{\lambda} &=
      6 \lambda^2
    - 3\lambda \bigg( 3 g_2^2 + \frac{3}{5} g_1^2 \bigg)
    + 3 g_2^4
    + \frac{3}{2} \bigg( \frac{3}{5} g_1^2 + g_2^2 \bigg)^2 \nonumber\\
   &\quad+ 4\lambda \, \tr \big[ Y_e^\dagger Y_e
                          + \hat{Y}_\nu^\dagger \hat{Y}_\nu
                          + \tilde Y_\nu^\dagger \tilde Y_\nu
                          + 3 Y_d^\dagger Y_d
                          + 3 Y_u^\dagger Y_u \big] \nonumber\\
   &\quad- 8 \,\tr\big[ Y_e^\dagger Y_e Y_e^\dagger Y_e
                 + \big( \hat{Y}_\nu^\dagger \hat{Y}_\nu
                       + \tilde Y_\nu^\dagger \tilde Y_\nu \big)
                   \big( \hat{Y}_\nu^\dagger \hat{Y}_\nu
                       + \tilde Y_\nu^\dagger \tilde Y_\nu \big) \nonumber \\ 
        &\qquad\ + 3 Y_d^\dagger Y_d Y_d^\dagger Y_d
                 + 3 Y_u^\dagger Y_u Y_u^\dagger Y_u \big]
\end{align}
Note that the new fields introduced in the inverse seesaw model
are all singlets under the SM gauge group. Therefore, the 
one-loop running of the $SU(3)_c\times SU(2)_L \times U(1)_Y$ gauge couplings is
described by the SM expressions
\begin{align}
  g^{-2}_1(\mu) &= g_1^{-2}(M_Z)
    - \frac{41}{80\pi^2} \log\bigg(\frac{\mu}{M_Z}\bigg) \,, \\
  g^{-2}_2(\mu) &= g_2^{-2}(M_Z)
    + \frac{19}{48\pi^2} \log\bigg(\frac{\mu}{M_Z}\bigg) \,, \\
  g^{-2}_3(\mu) &= g_3^{-2}(M_Z)
    + \frac{7}{8\pi^2} \log\bigg(\frac{\mu}{M_Z}\bigg)\,.
  \label{eq:ISSbetalast}
\end{align}

%------------------------------------------------------------------------------
\subsection{Beta functions for the type-I seesaw model}
\label{sec:type-I-from-iss}
%------------------------------------------------------------------------------

By comparing the mass matrix of the inverse seesaw model, \cref{eq:M-nu-tilde},
to the mass matrix of the type-I seesaw model
\begin{align}
  M_{\nu,\text{type-I}} \equiv
    \begin{pmatrix}
      0                  & Y^T_\nu v / \sqrt{2} \\
      Y_\nu v / \sqrt{2} & M \\
    \end{pmatrix} \,,
\end{align}
we can immediately read off that the RGEs for the type-I seesaw model are
obtained from \crefrange{eq:ISSbetafirst}{eq:ISSbetalast} by setting
$\hat{Y}_\nu = Y_\nu$, $M_N = M$, $\tilde{Y}_\nu = 0$, $m_D = 0$, and $\hat{m}_s = 0$.

%==============================================================================
\section{\texorpdfstring{Beta functions in the inverse seesaw model
         with gauged $B-L$}
         {Beta functions in the inverse seesaw model with gauged B-L}}
\label{sec:iss-B-L}
%==============================================================================

For the model with the gauged $U(1)_{B-L}$ symmetry, described in \cref{sec:ISSBL},
we find the following evolution equations:
\begin{align}
  16 \pi^2 \beta_{Y_\nu} &=
      Y_\nu \bigg[ \frac{1}{2}\lambda_{S}^T\lambda_{S}^{*} \bigg]
    + Y_\nu \bigg[ \frac{3}{2} Y_\nu^{\dagger} Y_\nu
    - \frac{3}{2} Y_e^\dagger Y_{e}
    + \tr (Y_e^\dagger Y_e)
    + \tr (Y_\nu^\dagger Y_\nu)  \nonumber\\
  &\qquad + 3\,\tr (Y_d^{\dagger} Y_d)
    + 3\,\tr ( Y_u^{\dagger} Y_u )
    - \frac{9}{20} g_1^2
    - \frac{9}{4} g_2^2
    - 6 g'^2 \bigg] \\[0.2cm]
  16 \pi^2 \beta_{Y_e} &= Y_e \bigg[
      \frac{3}{2} Y_e^\dagger Y_e
    - \frac{3}{2} Y_\nu^\dagger Y_\nu
    + \tr (Y_e^\dagger Y_e)
    + \tr (Y_\nu^\dagger Y_\nu )
    + 3\,\tr( Y_d^\dagger Y_d) \nonumber\\
  &\qquad
    + 3\,\tr( Y_u^\dagger Y_u )
    - \frac{9}{4} g_1^2
    - \frac{9}{4} g_2^2
    - 6 g'^2 \bigg] \\[0.2cm]
  16 \pi^2 \beta_{\lambda_{S}} &= \lambda_S \bigg[
      \tr(\lambda_S^\dagger \lambda_S)
    + Y_\nu^\ast Y_\nu^T
    + \lambda_S^\dagger \lambda_S
  - 15 g'^2 \bigg] \\[0.2cm]
  16 \pi^2 \beta_{Y_{u}} &= Y_u \bigg[
      \frac{3}{2} Y_u^{\dagger} Y_u
    - \frac{3}{2} Y_d^{\dagger} Y_d
    - \frac{17}{20} g_1^2
    - \frac{9}{4} g_2^2
    - 8 g_3^2
    - \frac{2}{3} g'^2 \nonumber\\
  &\qquad \vphantom{\frac{3}{2}}
    + \tr(Y_e^\dagger Y_e)
    + \tr(Y_\nu^\dagger Y_\nu)
    + 3\,\tr(Y_d^\dagger Y_d)
    + 3\,\tr(Y_u^\dagger Y_u) \bigg] \\[0.2cm]
  16 \pi^2 \beta_{Y_{d}} &= Y_d \bigg[
      \frac{3}{2} Y_d^\dagger Y_d
    - \frac{3}{2} Y_u^\dagger Y_u
    - \frac{1}{4} g_1^2
    - \frac{9}{4} g_2^2
    - 8 g_3^2
    - \frac{2}{3} g'^2 \nonumber\\
  &\qquad \vphantom{\frac{3}{2}}
    + \tr(Y_e^\dagger Y_e)
    + \tr(Y_\nu^\dagger Y_\nu)
    + 3\,\tr(Y_d^\dagger Y_d)
  + 3\,\tr(Y_u^\dagger Y_u) \bigg] \\
  16 \pi^2 \beta_{g'} &=  C g'^3
  \label{eq:beta-g'}
\end{align}
The numerical coefficient $C$ in the beta function for the $U(1)_{B-L}$ coupling
constant $g^\prime$ in the last line, \cref{eq:beta-g'}, depends on the number of
fields in the model and is given by
\begin{equation}
 C = \frac{2}{3} \sum_\text{fermions} Q_{B-L}^2
   + \frac{1}{3} \sum_\text{scalars} Q_{B-L}^2 \,.
\end{equation}
In our specific model, including four generations of $S$ and $S'$ each, we find
$C = \frac{97}{3}$.  For only three generations of $S$ and $S'$ fields, we
would have found $C = 27$ instead. The running of the quartic Higgs coupling
$\lambda$ and of the SM gauge couplings is the same as in the phenomenological
inverse seesaw model at one-loop level, see previous section. Note that we
neglect the running of all parameters in the scalar sector other than the
quartic Higgs coupling.  The running of these couplings influences the
evolution of the mixing parameters only indirectly by modifying the
$U(1)_{B-L}$ breaking scale.

Where possible, we have checked that we reproduce the applicable terms in the
renormalization group equations presented in ref.~\cite{Basso:2010jm} for a
$U(1)_{B-L}$ extension of the SM with three right-handed neutrinos, but without
an inverse seesaw mechanism.

% Bibliography
\bibliographystyle{JHEP}
\bibliography{ref.bib}
\end{document}